\documentclass[aps,twocolumn,superscriptaddress]{revtex4-1}
\usepackage{graphicx}
\usepackage{xcolor}
\usepackage[english]{babel}
\usepackage{amsmath}
\usepackage{amssymb}
\usepackage{bm}
\usepackage{makecell}
\usepackage{array}

\usepackage{float}

\usepackage{algorithm,algpseudocode}
\usepackage{wrapfig}

\definecolor{darkgreen}{RGB}{0, 150, 0}

\def\1{{\mathds{1}}}

\begin{document}

\author{Alice E. A. Allen}
\affiliation%[Center for Nonlinear Studies]
{Center for Nonlinear Studies, Los Alamos National Laboratory, Los Alamos, New Mexico 87545, United States}
\affiliation%[Theoretical Division]
{Theoretical Division, Los Alamos National Laboratory, Los Alamos, New Mexico 87545, United States}

\author{Nicholas Lubbers}
\affiliation%[CCS]
{Computer, Computational, and Statistical Sciences Division, Los Alamos National Laboratory, Los Alamos, New Mexico 87545, United States}
\author{Sakib Matin}
\affiliation%[Center for Nonlinear Studies]
{Center for Nonlinear Studies, Los Alamos National Laboratory, Los Alamos, New Mexico 87545, United States}
\affiliation%[Theoretical Division]
{Theoretical Division, Los Alamos National Laboratory, Los Alamos, New Mexico 87545, United States}
\author{Justin Smith}
\affiliation%[NVIDIA]
{Nvidia Corporation, Santa Clara, CA 9505, United States}
\author{Richard Messerly}
\affiliation%[Theoretical Division]
{Theoretical Division, Los Alamos National Laboratory, Los Alamos, New Mexico 87545, United States}
\author{Sergei Tretiak}
\affiliation%[Center for Nonlinear Studies]
{Center for Integrated Nanotechnologies, Los Alamos National Laboratory, Los Alamos, New Mexico 87545, United States}
\affiliation%[Theoretical Division]
{Theoretical Division, Los Alamos National Laboratory, Los Alamos, New Mexico 87545, United States}
\author{Kipton Barros}
\affiliation%[Center for Nonlinear Studies]
{Center for Nonlinear Studies, Los Alamos National Laboratory, Los Alamos, New Mexico 87545, United States}
\affiliation%[Theoretical Division]
{Theoretical Division, Los Alamos National Laboratory, Los Alamos, New Mexico 87545, United States}

%\iffalse

\title[]{Learning Together: Towards foundational models for machine learning interatomic potentials with meta-learning}
 
\begin{abstract}
The development of machine learning models has led to an abundance of datasets containing quantum mechanical (QM) calculations for molecular and material systems. However, traditional training methods for machine learning models are unable to leverage the plethora of data available as they require that each dataset be generated using the same QM method. Taking machine learning interatomic potentials (MLIPs) as an example, we show that meta-learning techniques, a recent advancement from the machine learning community, can be used to fit multiple levels of QM theory in the same training process. Meta-learning changes the training procedure to learn a representation that can be easily re-trained to new tasks with small amounts of data. We then demonstrate that meta-learning enables simultaneously training to multiple large organic molecule datasets.  As a proof of concept, we examine the performance of a MLIP refit to a small drug-like molecule and show that pre-training potentials to multiple levels of theory with meta-learning improves performance. This difference in performance can be seen both in the reduced error and in the improved smoothness of the potential energy surface produced.  We therefore show that meta-learning can utilize existing datasets with inconsistent QM levels of theory to produce models that are better at specializing to new datasets. This opens new routes for creating pre-trained, foundational models for interatomic potentials.
\end{abstract}

\maketitle

%\clearpage
\section{Introduction}

\begin{figure*}[htb!]
  \begin{center}
    \includegraphics[width=0.99\textwidth]{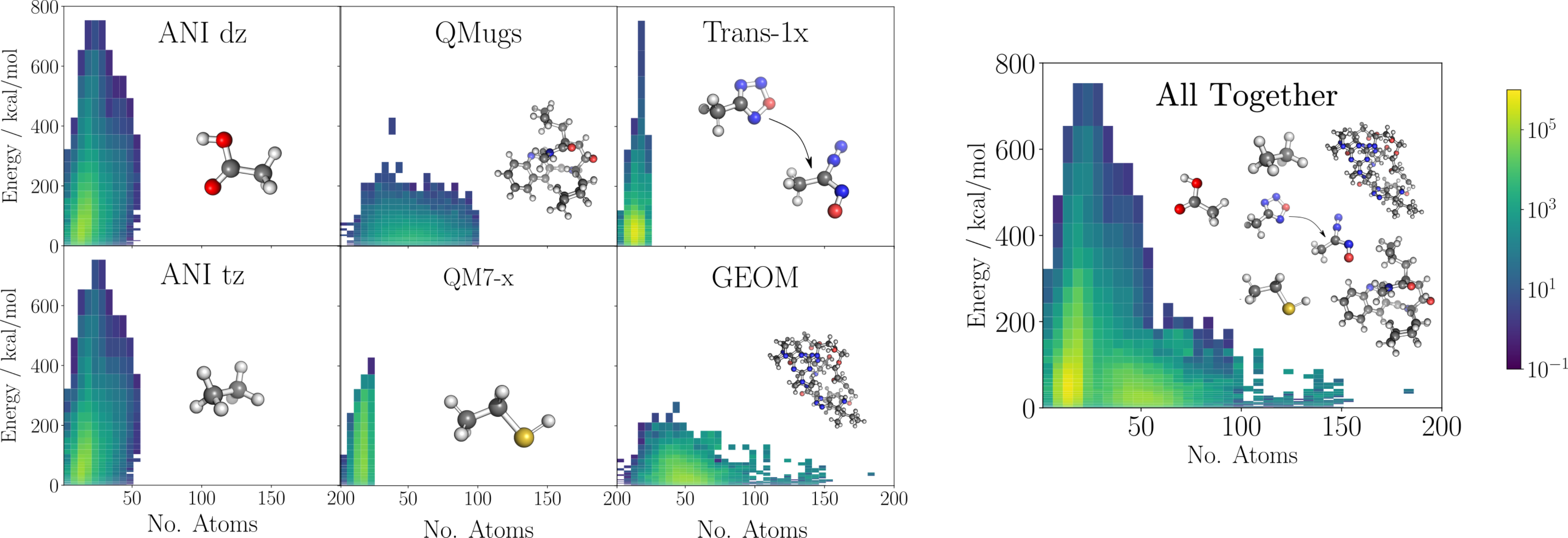}
  \end{center}
  \caption{A diverse collection of datasets, with varying levels of theory, molecule sizes, and energies, will be incorporated into a single meta-learned potential. The distributions of the number of atoms and energy of the structures contained in the datasets used for training a potential in this work are shown. The structures included contain only C,H,N,O. Energies are made comparable using linear scaling as detailed in Sec.~\ref{sec:multiple_org_mol}. }
  \label{fig:dataset_dis}
\end{figure*}

Machine learning is fundamentally changing and expanding our capabilities for modeling chemical and materials systems~\cite{ Isayev2017, Xie2018, Mikulskis2019, Pilania2021, Ouyang2018, Jha2018, Nandy2021, Keith2021}. A growing array of properties have been successfully predicted with machine learning models from materials' band gaps and formation energies to molecular energies and bond orders~\cite{Zhuo2018,Smith2017, Faber2016, Magedov2021}.  The development of machine learning models for various applications has involved the creation of a large number of datasets containing  quantum-mechanical calculations at different fidelities (levels of theory)~\cite{Nandi2022,Smith2020_1x_1ccx,Axelrod2022-wm,Hoja2021,Isert2022,Schreiner2022}. However, incorporating this multi-fidelity information into machine learning models remains challenging. In this work, we show that multiple datasets \emph{can} be used to fit a machine learning model, even if the datasets were calculated with many varying QM levels of theory. To overcome this challenge, we incorporate meta-learning techniques into the training process and subsequently demonstrate improvements in accuracy for multiple applications. The aim of meta-learning is to use a wide collection of data to train a machine learning model that can then be easily re-trained to specialized tasks and we demonstrate the applicability of the meta-learning method to MLIPs.

In the landscape of broader efforts to incorporate machine learning and molecular and material modelling, a particular attention has been paid to MLIPs~\cite{Smith2017, Schutt2017, bartok2010gaussian, Behler:2011it, ChemRev-MLFF,Mueller2020,Bartok:2018ih}. Accurate atomistic simulations rely on interatomic potentials that closely recreate the interactions present between atoms and molecules~\cite{Noe2020,Frenkel2001}. Recreating these interactions involves a trade-off between accuracy and computational cost, with quantum mechanical techniques  offering highly accurate simulations whilst classical force fields are fast and capable of modelling much larger systems over long timescales~\cite{Weiner1984,Jorgensen1988,Bartlett2007}. Within the last decade, MLIPs have increasingly been seen as a method that could provide a model that is both fast and accurate~\cite{Smith2017, Schutt2017, bartok2010gaussian}. However, the development of MLIPs that are transferable to unseen organic molecules requires datasets that cover a large fraction of chemical space. This requirement has lead to the production of numerous datasets~\cite{Nandi2022,Smith2020_1x_1ccx,Axelrod2022-wm,Hoja2021,Isert2022,Schreiner2022}. These datasets contain the quantum mechanical (QM) energies and forces of millions of structures spanning large regions of chemical space. However, the QM methods used  to calculate the energies and forces vary considerably. As different QM methods result in different potential energy surfaces, this inconsistency in QM techniques limits the extent that datasets can used together to fit potentials.

Numerous organic molecule datasets have been created for training MLIPs~\cite{Nandi2022,Smith2020_1x_1ccx,Axelrod2022-wm,Hoja2021,Isert2022,Schreiner2022}. However, a consensus on the best QM techniques to employ to create these datasets has never been reached as a compromise between accuracy and computational cost must always be considered when performing QM calculations. This lack of consensus has led to a variety of different software, methods, basis sets and exchange-correlation functionals being used. For example, the QM7-x and ANI-1x datasets both contain energies and forces for millions of small organic molecules. However, QM7-x was calculated using the PBE0 exchange-correlation functional with many body dispersion whilst ANI-1x was calculated with the $\omega$B97x functional and 6-31G* basis set~\cite{Smith2020_1x_1ccx,Hoja2021} and does not include dispersion effects. Therefore, these two datasets describe similar, but slightly different potential energy surfaces. If both datasets were joined together to train a potential then problems would likely arise as contradictory information is present. For example, identical structures at different levels of theory can have different energy and forces.  Whilst datasets from different sources have been fit together without further refinement~\cite{Tran2022}, this approach does not account for differences in the interactions described. Techniques exist in the machine learning literature to address the difference in the potential energy surface. 

Previous work on fitting MLIPs to multiple datasets is limited. In Ref.~\citenum{Smith2019}, a transferable molecular potential was first trained to $\sim$ 5 million density functional theory (DFT) training points before being refit, with frozen parameters, to 0.5 million CCSD(T)* energies. This technique, known as transfer learning has been used in several works~\cite{Chen2022,Taylor2009, Pan2010, Zaverkin2023, Kovacs2021}. The advantage of using transfer learning for training MLIPs is that it requires fewer calculations at a higher, and more expensive, level of theory. However, this kind of transfer learning technique, freezing neural network (NN) parameters, is limited to just two datasets. If we want to use multiple existing datasets, and expand the size and variety of training data, then new methods must be found. 

Fortunately, this problem is being explored in a branch of machine learning research known as meta-learning~\cite{Finn2017, Hospedales2002,Nichol2018,Huisman2021}. Meta-learning seeks to build a model that, although not specialized to any particular task, can be quickly re-trained to many new tasks - where a task is a specific learning problem. Furthermore, this retraining can be effective even if the amount of new data is limited~\cite{Finn2017,Nichol2018}.

\begin{figure*}[htb]
  \begin{center}
    \includegraphics[width=0.980\textwidth]{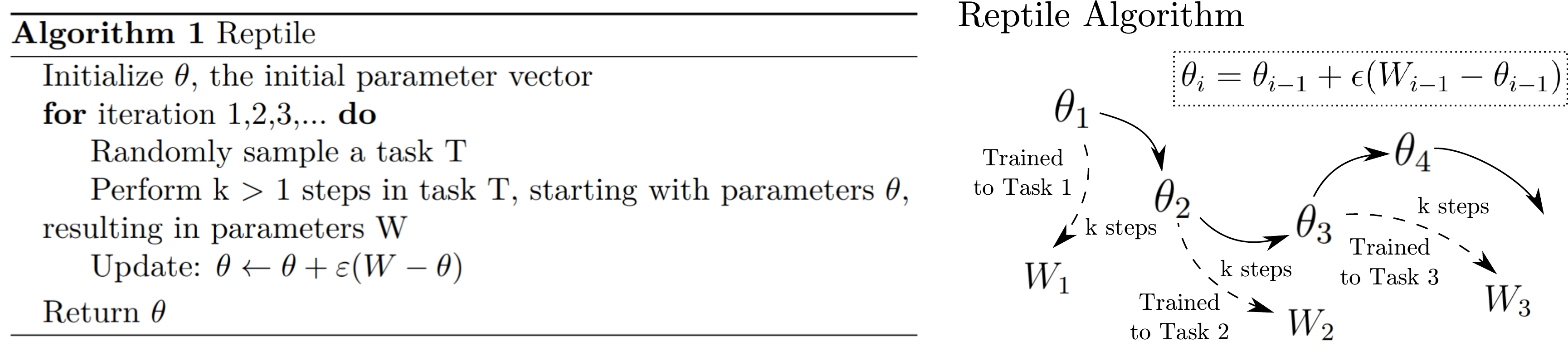}
  \end{center}
  \caption{This work uses Reptile to build a potential that incorporates information from multiple molecular datasets, calculated at different levels of theory. This meta-learned potential adapts well to new tasks, and outperforms potentials that were trained only to the data for a single task.}
  \label{fig:reptile_pic}
\end{figure*}

For transferable MLIPs, the concepts of tasks naturally lends itself to quantum mechanical datasets calculated with different methods. By using meta-learning techniques, we will show how information from multiple levels of theory can be incorporated together. We begin by investigating training data with multiple levels of theory for an individual aspirin molecule and for the QM9 dataset (which contains over 100,000 molecules in their equilibrium configuration). With these systems, the problems associated with naively combining datasets together are seen and the benefits of meta-learning are clearly observed in the test set errors.  We then move on to combining several large molecule datasets to pre-train an MLIP. Combining large organic datasets to fit MLIPs has never previously been attempted. Subsets, chosen using active learning, of six existing datasets (ANI-1x, GEOM, QMugs, QM7-x, Transition-1x and the QM9 dataset from Ref.~\citenum{Nandi2022}) were used to fit an adaptable potential using meta-learning -- see Fig.~\ref{fig:dataset_dis} for a visualization of the space the datasets cover~\cite{Nandi2022,Smith2020_1x_1ccx,Axelrod2022-wm,Hoja2021,Isert2022,Schreiner2022}. Figure~\ref{fig:dataset_dis} demonstrates the increase in chemical space possible when multiple datasets are combined together.  The benefits of pre-training are then shown by retraining to the 3BPA molecule and testing various properties. These tests show that pre-training models using meta-learning produces a more accurate and smoother potential. The benefits of pre-training include enhanced accuracy and generalization capabilities in modeling interatomic potentials. 

%In total, these datasets contain close to 90 million calculations, with over 200 different levels of theory. Large molecules with up to 100 heavy atoms are included and multiple conformer generation methods are employed. Additionally, dispersion effects are also incorporated into the potential and transition paths for over 10,000 reactions are present. Access to this breadth of training data for transferable organic potentials is unprecedented and only possible due to integration of meta-learning techniques into training. By incorporating such a large range of information, and using meta-learning techniques, we will show that 

Training machine learning models to large amounts of data before re-training to a specific task is  related to the concept of foundational models~\cite{Bommasani2022}. This concept has been used to create large language models, ie. GPT-4, which have been pre-trained to extremely large datasets before being fine-tuned to specific tasks, i.e. ChatGPT which is fine-tuned for conversational usage~\cite{Openai2023}. Creating foundational models allows a wide range of information to be encoded before specialisation. With meta-learning techniques, we can now pre-train interatomic potentials to numerous large datasets and this is a step towards foundational models for MLIPs -- MLIPs that could be quickly re-trained to diverse molecular systems.

The number of QM datasets has grown rapidly over the last few years. However, a major bottleneck in exploiting this information has been the absence of methods that can effectively combine all of this information.  In this work, we have overcome this limitation by exploiting techniques which enable the incorporation of datasets with different fidelities. Whilst we focus on MLIPs, these techniques are applicable to the wide range of predictive models that exist for material and molecular property prediction. By showing how meta-learning can be applied, we aim to encourage researchers to fully utilize the vast amount of existing data that the scientific community has already collected.

\section{Methods}

\subsection{Meta-Learning Algorithm}

Meta-learning is an area of machine learning concerned with improving the learning process to produce models that can easily adapt to new problems~\cite{Finn2017, Hospedales2002,Nichol2018,Huisman2021}. A key component of meta-learning is the concept of different `tasks'. Tasks are datasets with similar properties but slight differences. For example, if we were interested in animal classification of a cat and a dog, a similar task might be to classify a lion and a bear. The task is not the same but we would expect fundamental similarities in the model needed to perform the classification. By using a meta-learning algorithm to learn multiple different tasks, less data will be required when a new learning problem is introduced.

The objective of meta-learning algorithms is to train a model that can generalize more easily to new data\cite{Hospedales2002, Huisman2021}. We will use meta-learning to fit multiple different QM datasets with slightly different properties. To our knowledge, meta-learning for MLIPs has not been previously carried out, although it has been used in other areas of science~\cite{Yangzesheng2021,Nie2021,WangJ2021}.

The meta-learning algorithm we have chosen to fit multiple datasets for MLIPs is called Reptile~\cite{Nichol2018}. Reptile works by repeatedly sampling a task (a dataset), performing a limited number of optimization steps on the task and then updating the weights of the machine learning model towards the new weights.  Reptile was chosen over other meta-learning algorithms such as MAML~\cite{Finn2017} as Reptile is simpler to implement and therefore more likely to be adopted by the wider community. A comparison of  methods such as MAML for interatomic potentials will therefore be left to future work.

Reptile is described in Algorithm \ref{fig:reptile_pic} with a visual illustration also given. The algorithm works by separating the training data into distinct learning problems (tasks). An individual task is selected and multiple optimization steps are performed. The parameters of the model are then updated. A new task is then selected and the procedure is repeated multiple times. This moves the model to a region of parameter space where it can readily move between the different datasets present. 

Throughout this work, the $k=1$ result is used as comparison point. This is because when $k=1$ the algorithm becomes equivalent to stochastic gradient descent on the expected loss over all the  training tasks~\cite{Nichol2018}. This is referred to as joint training in Ref.~\citenum{Nichol2018} At $k=1$, the algorithm is not expected to account for differences in the QM theory but still uses all the information present from the datasets.

\subsection{Interatomic Potential}
In this work, we have used the NN architecture implemented in torchANI with the same structure as the ANI-1x model~\cite{Smith2017, Smith2019}. However, the meta-learning techniques described are not specific to this form of model and there is no reason that they could not be applied to other machine learning models that employ similar iterative solvers. 

The hyperparameters used for the ANI potential are the same as those used for previous training to the ANI-1x and ANI-1ccx datasets, see Ref.~\citenum{Smith2019} for more details. 

\subsection{Datasets}

\begin{table*}[htb]
\begin{tabular}{|c|c|c|c|c|c|c|c|}
\hline
\textbf{Dataset} & \textbf{\begin{tabular}[c]{@{}c@{}}Unique \\ Compounds\end{tabular}} & \textbf{\begin{tabular}[c]{@{}c@{}}Total \\ Conformers\end{tabular}} & \textbf{\begin{tabular}[c]{@{}c@{}}Heavy \\ Atoms\\  Max\end{tabular}} & \textbf{\begin{tabular}[c]{@{}c@{}}Conformer\\  Generation\end{tabular}} & \textbf{Method} & \textbf{Dispersion} & \textbf{\begin{tabular}[c]{@{}c@{}}Transition\\  Paths\end{tabular}} \\ \hline
\textbf{QM9} & 133,885 & 133,885 & 9 & None & \begin{tabular}[c]{@{}c@{}}76 DFT \\ Functionals\end{tabular} & Yes & No \\ \hline
\textbf{ANI-1x} & $\sim$64,000 & \begin{tabular}[c]{@{}c@{}}4,956,005\\ and \\ 4,617,229\end{tabular} & 8 & \begin{tabular}[c]{@{}c@{}}Normal Mode Sampling,\\  MD sampling,\\  Torsional Sampling,\\  Active Learning \end{tabular} & \begin{tabular}[c]{@{}c@{}}$\omega$B97x/\\ 6-31G* \\ and\\  $\omega$B97x/\\ def2-TZVPP\end{tabular} & No & No \\ \hline
\textbf{QMugs} & 665,911 & 1,992,984 & 100 & \begin{tabular}[c]{@{}c@{}}Meta Dynamics \\ with xTB\end{tabular} & \begin{tabular}[c]{@{}c@{}}$\omega$B97X-D/\\ def2-SVP\end{tabular} & Yes & No \\ \hline
\textbf{GEOM} & 437,724 & 32,657,609 & 91 & \begin{tabular}[c]{@{}c@{}}Meta Dynamics \\ with CREST\end{tabular} & \begin{tabular}[c]{@{}c@{}}r2scan-3c/\\ mTZVPP\end{tabular} & Yes & No \\ \hline
\textbf{QM7-x} & 41,537*  & 4,195,237 & 7 & \begin{tabular}[c]{@{}c@{}}Normal-mode\\  Sampling with DFTB\end{tabular} & PBE0+MBD & Yes & No \\ \hline
\textbf{Transition-1x} & 10,073 Reactions & 9,644,740 & 7 & Nudged Elastic Band & \begin{tabular}[c]{@{}c@{}}$\omega$B97x/\\ 6-31G(d)\end{tabular} & No & Yes \\ \hline
\textbf{ANI-1ccx} & $\sim$64,000 & 489,571 & 8 & Active Learning from ANI-1x & \begin{tabular}[c]{@{}c@{}}CCSD(T)*/\\ CBS\end{tabular} & Yes & No \\ \hline
\end{tabular}
\caption{The properties of the datasets used in this work are shown in the table. (*including conformational isomers)}
\end{table*}

\subsubsection{Aspirin}

Aspirin structures were produced by molecular dynamic simulations at 300K, 600K and 900K. Density Functional based Tight Binding (DFTB) was used to perform the MD simulations and a total of 400 structures were created for each temperature. QM calculations of the energies and forces were then performed on these structures with three levels of theory: DFT with the  $\omega$B97x exchange-correlation function and 6-31G* basis set, DFT with Becke, 3-parameter, Lee–Yang–Parr (B3LYP) exchange-correlation functions and def2-TZVP basis set and Hartree-Fock with the def2-SVP basis set for 300K, 600K and 900K respectively. These datasets were used to pre-train a molecular potential. The pre-trained potential was then refit to a new dataset of MD configuration at the M{\o}ller–Plesset (MP2) level of theory with the def2-SVP basis set (a more accurate level of theory). The training dataset for refitting used 400 MD configurations sampled at 300K whilst the test set contained structures at 300K,600K and 900K. A batch size of 8 was used for training. 

\subsubsection{QM9}
The QM9 dataset contains over 100,000 equilibrium structures for small organic molecules with up to 9 heavy atoms~\cite{Ramakrishnan2014}. In Ref.~\citenum{Nandi2022}, the QM9 dataset was recalculated with 76 different exchange-correlation functionals and 3 basis sets~\cite{Nandi2022}.

\subsubsection{Multiple Organic Molecules}
\label{sec:multiple_org_mol}
Seven separate datasets were chosen to fit a potential to organic molecule potential that could be easily re-trained to new data. The seven datasets used for meta-learning were chosen to cover both diverse regions of chemical space and multiples levels of theory -- including the accurate recreation of dispersion effects. The chemical space covered included reactive paths and biologically and pharmacologically relevant structures. Whilst ANI-1x does cover a large number of conformations for organic molecules, it has limitations. This is demonstrated by Fig.~\ref{fig:dataset_dis} and Fig.~S1. Figure \ref{fig:dataset_dis} demonstrates how the additional datasets increase the size of the molecules and range of energies included. The $E_0$ energy is calculated using linear fitting an then subtracted from each dataset. The minimum energy for each dataset is then shifted to zero.  Whilst it is not covered in this work as we use the ANI potential, including larger molecules in datasets may be increasingly important for newer generations of interatomic potentials that include message passing and describe longer length scales~\cite{Batzner2022,Lubbers2018}. Figure~S1 shows the distribution of uncertainty for the ANI-1x potential across the dataset space. Whilst ANI-1x dz, ANI-1x tz, GEOM and QMugs have similar probability distributions, QM7-x and Transition-1x contain larger uncertainties. Transition-1x contains reactive structures that are not contained in the original dataset and therefore higher uncertainties are expected. For QM7-x, there are also higher uncertainties and this may be due to the different sampling techniques used. 

A property that is not shown in Table~1 is the software used for the DFT calculations. Even when the same level of theory is used, we can expect different software to give slightly different results. This will cause further discrepancies between the datasets as a variety of codes are employed. For example, although Transition-1x and ANI-1x are calculated at the same level of theory, Transition-1x is calculated with the ORCA program whilst ANI-1x is calculated with Gaussian~\cite{Neese2020,g16}. 

%Additionally, there is not a standard format for the datasets. This meant that a pre-processing step was required prior to fitting. This was not trivial and emphasises the need for a standardized format. 

The individual description and justification for including each dataset used is as follows: 
\begin{itemize}
\item QM9 - This dataset contains a diverse range of 76 functionals and 3 basis sets for small equilibrium organic molecules~\cite{Nandi2022}.
\item ANI-1x - This is a large dataset of small (up to 8 heavy atoms) organic molecules generated with active learning methods~\cite{Smith2020_1x_1ccx}.
\item QMugs - This dataset includes the largest molecules with up to 100 heavy atoms. It specializes in including drug-like molecules~\cite{Isert2022}
\item GEOM - This is the largest dataset and contains both large molecules and drug-like molecules~\cite{Axelrod2022-wm}.
\item QM7-x - This is also a large dataset of small (up to 7 heavy atoms) organic molecules but has dispersion accurately described with many-body dispersion~\cite{Hoja2021}
\item Transition-1x - This datasets includes minimum energy paths for 12,000 reactions~\cite{Schreiner2022}.
\item ANI-1ccx - This dataset contains coupled cluster level theory calculations for a subset of the ANI-1x dataset~\cite{Smith2020_1x_1ccx}.
\end{itemize}

Other datasets considered for inclusion include SPICE, PubChemQC-PM6 and Tensormol~\cite{Eastman2022,Nakata2020,Yao2018}. However, with the existing datasets a sufficient representation of chemical space is covered. It is also worth noting that retraining to recreate the specific properties of the excluded datasets would also be quickly possible with the meta-learning potential. 

\subsection{Meta-learning Hyperparameter Optimization}
There are three parameters in the Reptile algorithm. These control the number of steps ($k$) taken at each optimization step, how the parameters are updated ($\epsilon$) from the task's individual NN parameters and the maximum number of epochs used for retraining. The number of epochs was investigated to see whether restricting the training improved accuracy by ensuring the potential remained close to the meta-learned potential or if longer retraining improved results. For a detailed discussion of the hyper parameters chosen when fitting to the seven separate datasets, see Section S1.2. The $\epsilon$ value used throughout this work is $\epsilon=1$ whilst the $k$ value is changed depending on the problem. The maximum number of epochs used for retraining for the meta-learning algorithm with $k>1$ is restricted to 150 epochs. 

\iffalse
\begin{figure}[h]
  \begin{center}
    \includegraphics[width=0.50\textwidth]{./Graphs/Stages_fitting.pdf}
  \end{center}
  \caption{The stages of fitting in the meta-learning process. *The QM9 dataset is the exception with only 10,000 structures. }
  \label{fig:stages_fitting}
\end{figure}
\fi

\subsection{Stages of Fitting for the Organic Molecule datasets}
 
In the first iteration, 100,000 structures were taken randomly from the ANI-1x, QMugs, GEOM, QM7-x and Transition-1x datasets. For QM9, 10,000 structures were used for each level of theory. This is restricted as 276 levels of theory exist, and each theory level samples different structures in the QM9 dataset. After the first iteration, the highest error structures were added to the next iteration~\cite{Smith2018-less}. The cutoffs used for adding structures are described in SI~1.6.  This process was repeated 3 times. A diagram of the process is show in Fig.~S3. 

\section{Results}

\subsection{A Simple Case Study on Aspirin}

As the initial test case we investigate the performance of meta-learning on a dataset containing a single aspirin molecule. Aspirin structures were produced by molecules dynamic simulations at 300K, 600K and 900K. The QM energies and forces were then calculated at three different levels of theory: two distinct DFT functionals, and Hartree-Fock. This created three different datasets, with each temperature corresponding to a different level of theory. These three datasets were used to pre-train a molecular potential to the energy and forces of 1,200 structures. The pre-trained potential was then refit to a new dataset of 400 MD configuration at the MP2 level of theory from the 300K simulation. 

\begin{figure}[h]
  \begin{center}
    \includegraphics[width=0.49\textwidth]{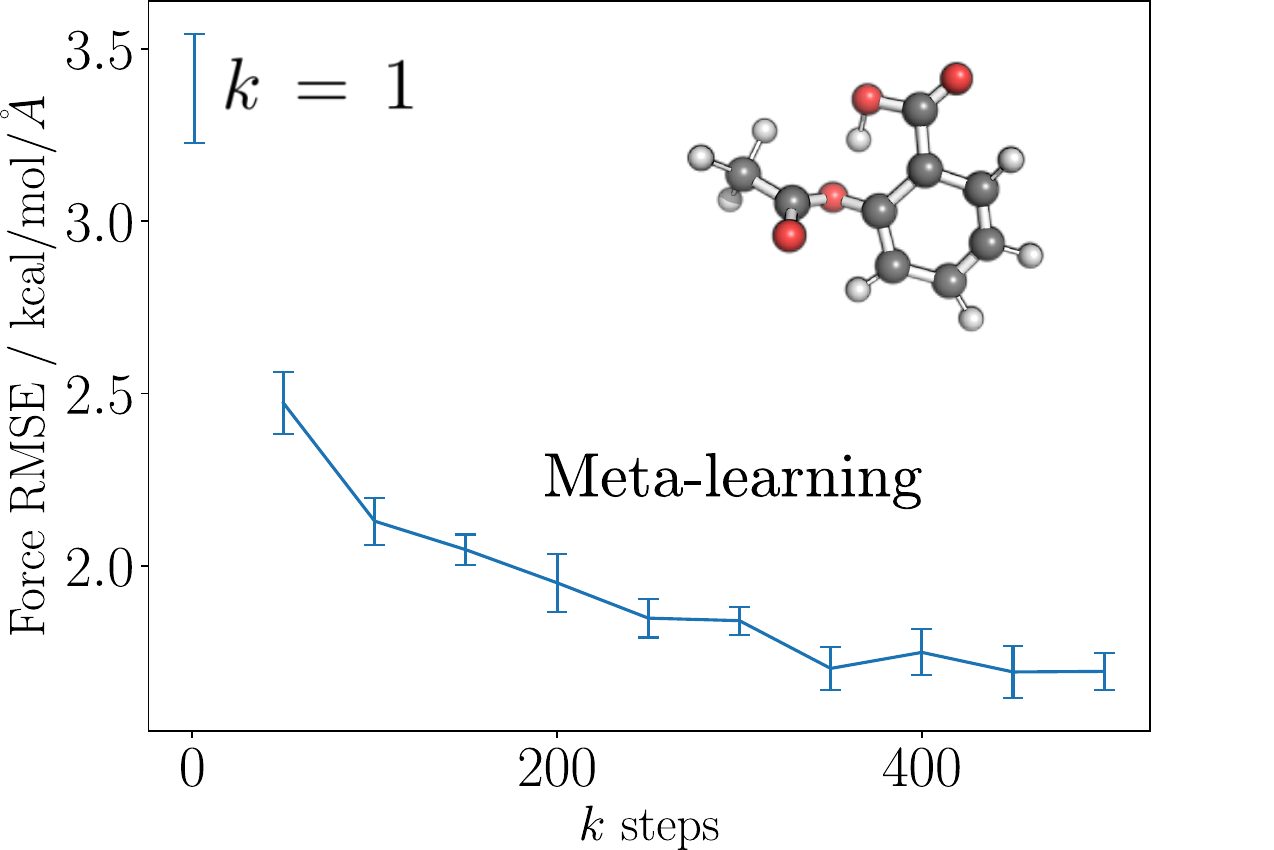}
  \end{center}
  \caption{The error as a function of the value of $k$ used in the meta-learning algorithm for an aspirin molecule. The potential is first pre-trained to multiple levels of theory before retraining to 400 structures at MP2 level of theory. When no pre-training is used the root mean squared error (RMSE) is 5.35 $\pm$ 0.41  kcal/mol/ \AA.}
  \label{fig:asp}
\end{figure}

The change in the RMSE error for the forces is shown with the value of $k$ used in the meta-learning algorithm in Fig.~\ref{fig:asp}. The $k$ parameter controls the number of steps taken towards each dataset. As $k$ is increased the speed of the algorithm also increases and this is an additional consideration in choosing the optimal value. In the limit of $k \to \infty $ the algorithm would correspond to iterative training to each dataset and then transfer learning to a new task. However, while this may work for small problems, this approach is impractical for large datasets. 

Figure~\ref{fig:asp} shows that as the $k$ parameter is increased the error in the test set decreases with the minimum error at around $k=400$. There is therefore an improvement in test set error in comparison to both no pre-training (5.35 $\pm$ 0.41 kcal/mol/ \AA) and $k=1$ (3.38 $\pm$ 0.16 kcal/mol/ \AA). Note that $k=1$ effectively corresponds to simultaneous training to all tasks.  Therefore, when we attempt to combine multiple datasets at different levels of theory an improvement in performance can be seen when meta-learning is incorporated into the training process.

\subsection{Meta-learning many levels of theory using QM9}

Next, we move onto the QM9 dataset that contains multiple different small organic molecules in their equilibrium structures. The QM9 dataset has been calculated at 228 different levels of theory and therefore provides an ideal dataset for analysing meta-learning techniques. We can use this dataset to test whether meta-learning can develop a potential which can be refit to a new level of theory encountered for the QM9 dataset with less data. In order to do this, a subset of the QM9 dataset was used to train a potential to 10,000 molecules, 50 different exchange-correlation functionals and three different basis set. The potential was then refit to a new exchange-correlation functional, that had not been previously encountered, and the performance of this new model was assessed and compared to no pre-training and $k=1$ meta-learning.
%Meta-learning with $k=1$ is when pre-training is carried out without allowing multiple optimization steps for each different dataset, thereby ignoring the difference in the levels of theory present. 

\begin{figure}[h!]
  \begin{center}
    \includegraphics[width=0.40\textwidth]{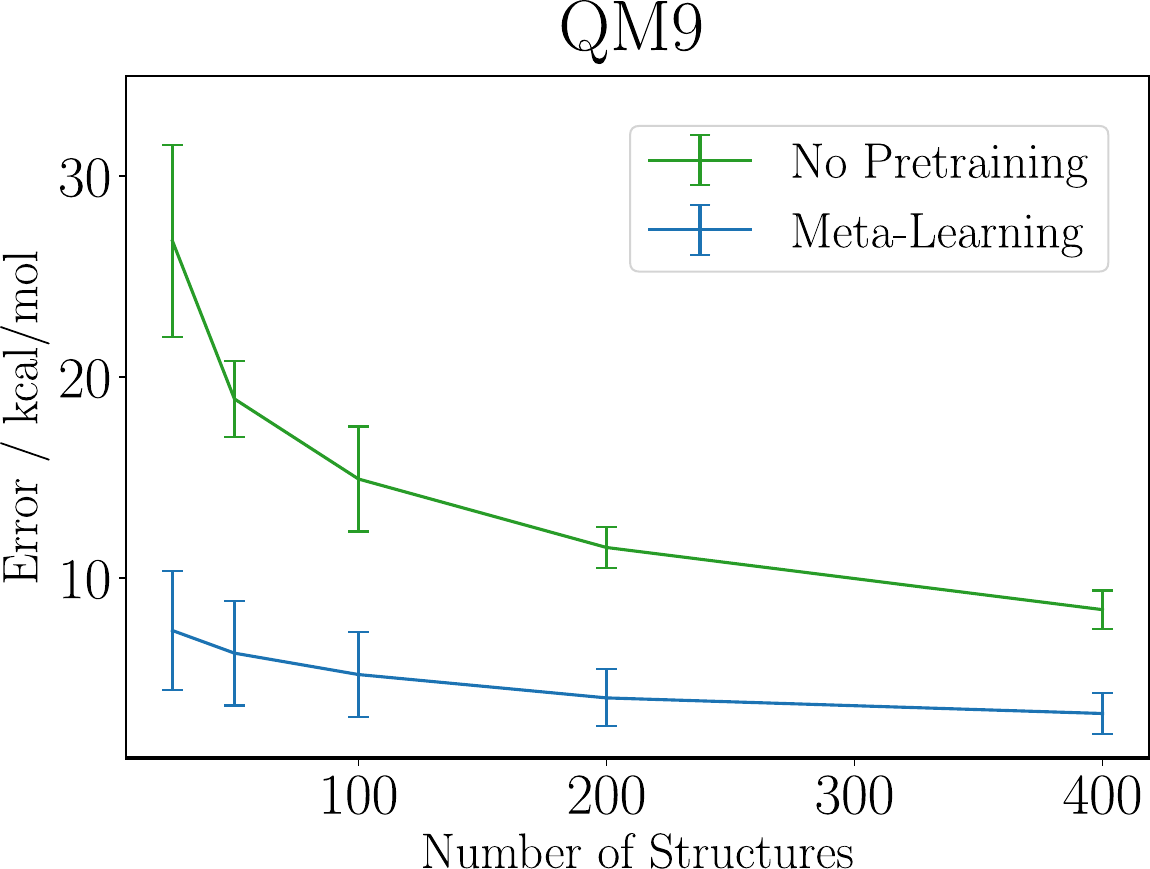}
  \end{center}
  \caption{The error as a function of the number of structures averaged across 5 different DFT functions and 3 basis sets, for different fitting procedures for the QM9 dataset.  The meta-learning algorithm fits the different levels of theory together with $k=10$ used. The standard deviation of the error bars corresponds to the variation across different, randomly selected, specialization tasks. Freezing the parameters before retraining was also attempted for 400 structures, however, no improvement in accuracy was observed.   }
  \label{fig:QM9}
\end{figure}

The test set error for the meta-learning potential refit to a new level of theory in the QM9 dataset is shown in Fig.~\ref{fig:QM9}. Pre-training the potential greatly improves the test set error for this case. In Fig.~S9 a comparison between meta-learning and $k=1$ is shown and we see that $k=1$ does not perform as well as $k=10$. This is because it does not account for the discrepancy in the interaction present. These results show that even when the number of levels of theory is relatively large, at 150, and multiple molecules are present that meta-learning improves test set error over $k=1$.

\subsection{Making the most of scarce data at CCSD(T) level}

\begin{figure}[htb!]
  \begin{center}
    \includegraphics[width=0.40\textwidth]{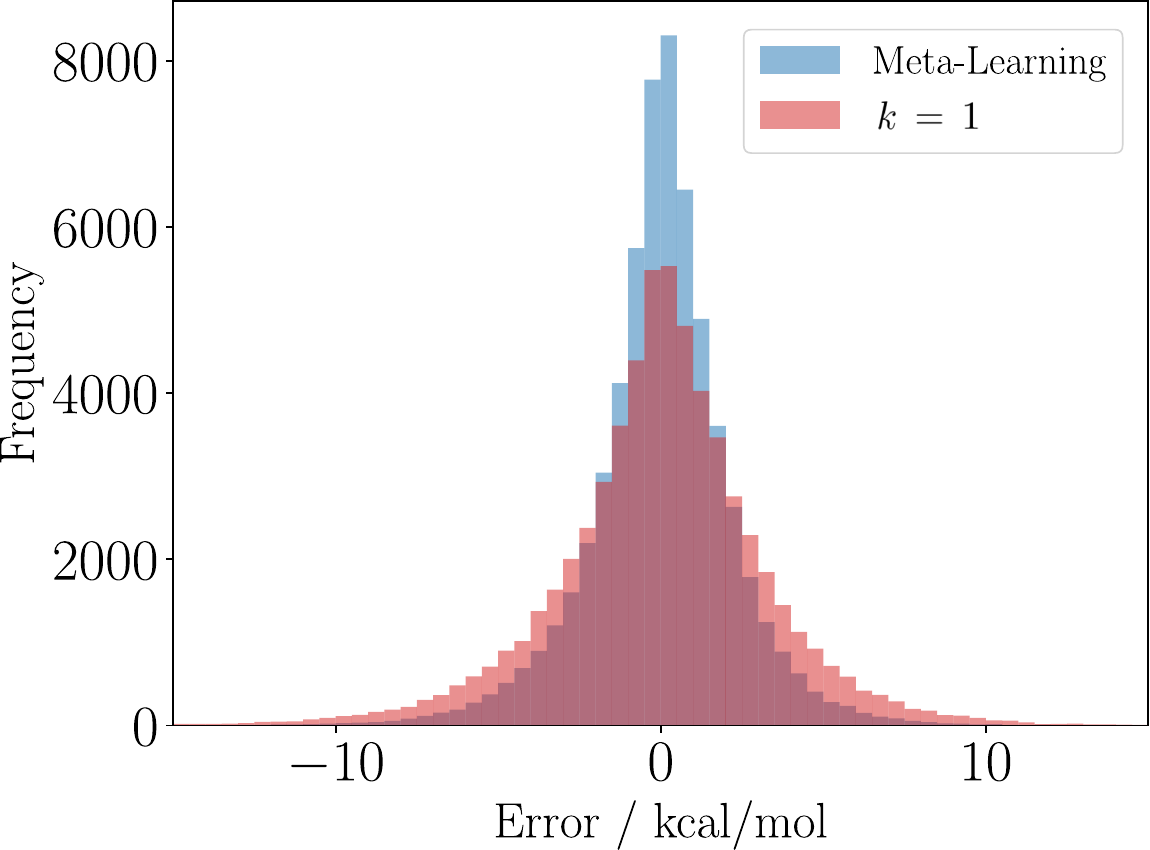}
  \end{center}
  \caption{The error distribution for the CCSD(T) specialization task after pre-training to ANI-1x with meta-learning ($k=50$) using and $k=1$. The RMSE is 3.47 kcal/mol and 2.39 kcal/mol. }
  \label{fig:ani_ccsdt_figs}
\end{figure}

We will now move to the datasets used to train transferable interatomic potentials. As a starting example, we will look at pre-training to the multiple levels of theory ($\omega$B97x/ 6-31G*  and $\omega$B97x/ def2-TZVPP) contained in the ANI-1x dataset~\cite{Smith2020_1x_1ccx}. We will then retrain to the ANI-1ccx dataset~\cite{Smith2020_1x_1ccx}. Figure~\ref{fig:ani_ccsdt_figs} shows the distribution in error when pre-training to multiple levels of theory with meta-learning and $k=1$. The RMSE is 3.30 $\pm$ 0.10 kcal/mol and 2.39 $\pm$ 0.00 kcal/mol for $k=1$ and meta-learning respectively. Therefore, we can again see that meta-learning with a higher $k$ values improves results compared to $k=1$. The comparative results for direct training to $\omega$B97x/ 6-31G*  and $\omega$B97x/ def2-TZVPP  and then transfer learning to CCSD(T) is 2.20$\pm$ 0.01 kcal/mol  and 2.09$\pm$0.02 kcal/mol respectively . Therefore, in this case fitting to multiple datasets does not improve results over fitting to just one. This is in part because both datasets contain the same structures and cover the same chemical and configurational space. The potential trained to multiple organic datasets was also refit to the CCSD(T) dataset and the benefits of meta-learning  over $k=1$ were also seen with errors of 2.89$\pm$ and 3.32$\pm$ respectively. However, this is notably higher than training to the ANI-1x dataset alone. The CCSD(T) dataset is a subset of the ANI-1x dataset and contains identical structures. For these cases, adding additional data in other areas of chemical space may not improve results. 

\begin{figure*}[htb!]
  \begin{center}
    \includegraphics[width=0.99\textwidth]{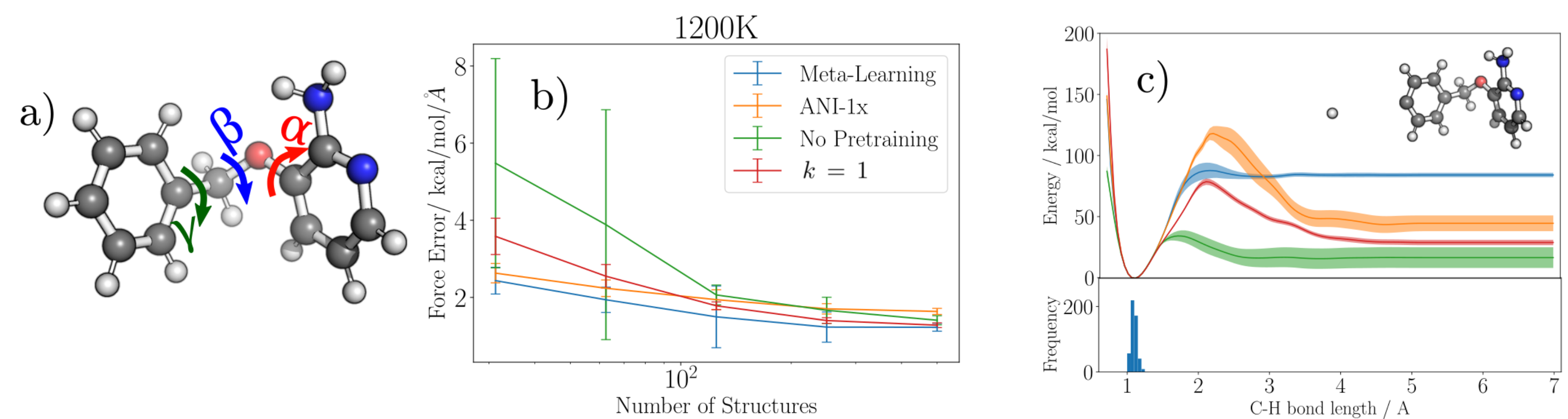}
  \end{center}
  \caption{The a) 3BPA molecule, b) the force error versus the number of structures used for the final training and c) hydrogen bond dissociation for 3BPA with the frequency of C--H bonds in the 3BPA training set. The results shown in part b) demonstrate the difference in performance with varying pre-training approaches -- with meta-learning producing the lowest energy results. The bond dissociation calculated using DFT is 119.6kcal/mol and 111.2kcal/mol with the bond dissociation energies estimator from Ref.~\citenum{John2020}. All other atoms are fixed with the removal of the hydrogen atom. The frequency of the C--H bond is for the 3BPA dataset used for re-training and the maximum length of the bond is 1.27\AA.  }
  \label{fig:3bpa_figs}
\end{figure*}

\subsection{Training to multiple transferable organic molecule datasets}

\begin{figure*}[htb!]
  \begin{center}
    \includegraphics[width=0.95\textwidth]{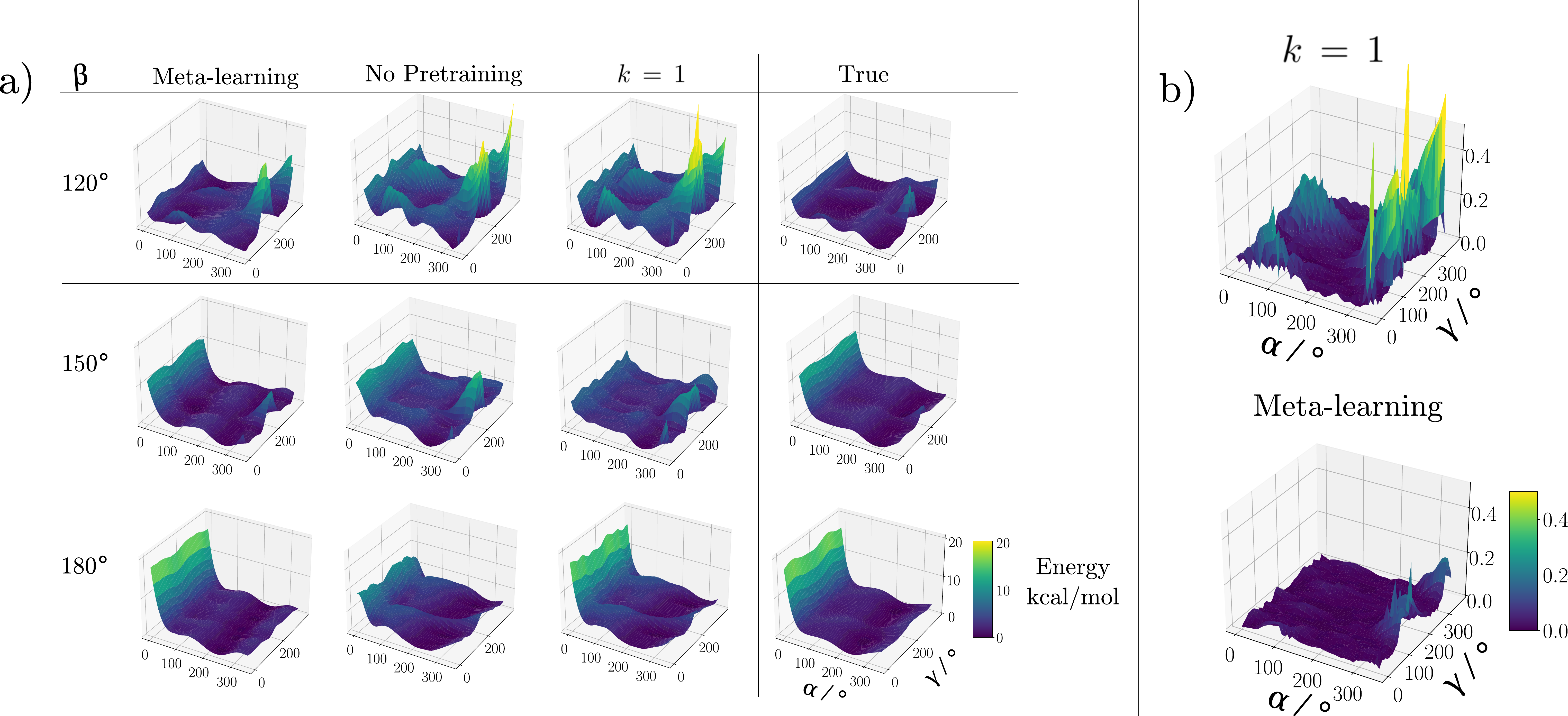}
  \end{center}
  \caption{a) Torsional energy scans (of $\alpha$ and $\gamma$) for potentials with different pre-training approaches. The $\beta$ angle is set to three different values. The potentials are re-trained to 62 structures from the 3BPA dataset. An ensemble of 8 models is used. Part b) shows the gradient of the torsional energy surface with respect to $\alpha$. The units are kcal/mol/degrees.   }
  \label{fig:dih_3BPA}
\end{figure*}

Numerous datasets have been created that contain quantum mechanical calculations for organic molecules. However, as these datasets use different levels of theory and software, combining the information from different datasets requires advanced training techniques. By using meta-learning, a pre-trained model was created that uses information from seven different datasets. This is the first instance, to our knowledge, of combining information from multiple organic molecule datasets in this manner.

We have already seen that meta-learning  can improve results compared to $k=1$ when multiple datasets are used. We will now use the pre-trained model to explore the benefits of pre-training with meta-learning in comparison to no pre-training, and $k=1$ when retraining to a single molecular system. The pre-trained model was re-trained to the 3BPA dataset taken from Ref.~\citenum{Kovacs2021} and various properties explored~\cite{Cole2019}. 

The first properties we will analyze are the energy and force RMSE errors. The force errors for a dataset taken from MD at 1200K is shown in Fig.~\ref{fig:3bpa_figs} with the energy and force learning curves for datasets at 300K, 600K and 1200K given in Fig.~S4. From these graphs, the improved performance of pre-training using the meta-learning approach (with three passes through the dataset) to both $k=1$ and no pre-training can be seen for energies and forces. Therefore, just by adapting the training scheme, with no change in the model architecture or the dataset itself, consistent improvements in accuracy can be seen with meta-learning. The importance of the training method used has previously been seen in Ref.~\citenum{Shao2021}. Here we see how it can improve performance for fitting multiple datasets together. In comparison to when the ANI-1x model is used for pre-training, meta-learning performs slightly better at force errors but slightly worse for energy predictions. Given that the ANI-1x model is fit to the same level of theory as the 3BPA dataset, the performance of the meta-learning potential is encouraging. 

However, it is known that RMSE errors alone are not enough to verify the performance of a potential~\cite{Fu2022,Kovacs2021}. We will therefore examine additional properties. The 3BPA molecule has three central dihedral angles which are illustrated in Fig.~\ref{fig:3bpa_figs}. The energy scans along these dihedral angles are shown in Fig.~\ref{fig:dih_3BPA} with the model refit to the energies and forces of just 62 3BPA conformations. When no pre-training is used, the surface at $\beta=120$ significantly over-estimates the high energy point and lacks smoothness. A similar shape is seen for the $k=1$ potential. However, when meta-learning is used for pre-training the surface remains noticeably smoother with significantly less over prediction. When $k=1$ is used, multiple different potential energy surfaces are combined together in a nonphysical way which destroys the smoothness of the underlying potential. The error in the gradient of the 2D energy surface is shown in Fig.~\ref{fig:dih_3BPA} b) and emphasizes this difference in smoothness. When meta-learning is used, the contradiction in the potential energy surface described is corrected resulting in a smoother model. When no pre-training or $k=1$ is used, an additional problem can occur with the high energy regions at $\alpha=0$ failing to be recreated for the $\beta=180$ and $\beta=150$ scan respectively. In contrast, both the meta-learning pre-training model correctly recreate this behaviour. The results for ANI-1x pre-training are given in Fig.~S6. 

One advantage of pre-training with multiple datasets over ANI-1x or QM7-x, is that reactive systems can be added that are not contained in ANI-1x. To test if this information has been effectively passed to the meta-learning potential, hydrogen bond dissociation for the 3BPA molecules was performed. There is no reactive information contained within the 3BPA training set and so this test relies entirely on the information contained in the pre-training. 

Figure \ref{fig:3bpa_figs} shows the change in energy as a hydrogen molecule is removed from the 3BPA. The potential pre-trained with meta-learning recreates the smooth dissociation curve expected. In contrast, when no pre-training, $k=1$ or ANI-1x is used the curve lacks smoothness and has an additional barrier present. In Fig.~S7, the bond dissociation energy when just 31 structures are used for retraining. Even in this low data limit the smooth dissociation curves for the meta-learning potential remain. To demonstrate that this is not unique to 3BPA, the hydrogen bond dissociation for ethanol is shown in Fig.~S8. Again, $k=1$ fails to recreate the smooth curve expected whilst the meta-learning potential captures the correct shape.

We have therefore shown how meta-learning can be used to combine multiple datasets and the resulting improvements in the error, torsion energy scans and bond dissociation. Joint-fitting can improve on no-pre-training. However, not accounting for the difference in QM level of theory causes a reduction in performance that can be seen in the test set errors, smoothness of the potential and performance in extrapolation regions.

\section{Conclusion}
%Summary
The quantum mechanical properties of millions of molecular species and many materials systems have already been calculated and composed into extended datasets~\cite{Nandi2022,Smith2020_1x_1ccx,Axelrod2022-wm,Hoja2021,Isert2022,Schreiner2022}. However, the varying levels of theory
used to perform the QM calculations has previously prevented different datasets being used together to make machine learning models, for example for MLIPs. In this work, we have shown that meta-learning techniques can be used to jointly fit multiple datasets and demonstrated the improvement in performance that results from including a diverse selection of datasets. 

We show the wide applicability of meta-learning by creating MLIPs for a variety of systems, from a single aspirin molecule to the ANI-1ccx dataset. By pre-training a model to multiple large organic molecule datasets we show that these datasets (QM7-x, QMugs, ANI-1x, Transition-1x and GEOM) can be combined together to pre-train a model. The benefits of using a pre-trained model are then shown for the 3BPA molecule, with a more accurate and smoother potential produced. Meta-learning greatly expands the variety of fitting data available for MLIPs and establishes the possibility of creating readily pre-trained, foundational models for MLIPs. 

%pre-training interatomic potentials
Pre-training machine learning models has been extensively discussed in the machine learning literature in recent years~\cite{Han2021,Hu2020,Hendrycks2019}. Whilst pre-training has been carried out for MLIPs, its use has been limited to training from one dataset to another~\cite{Smith2019, Kovacs2021, Chen2022}. With techniques such as meta-learning, this pre-training does not need to be limited to one specific dataset but can include large numbers of existing datasets. In this work, we added only a single reactive dataset to pre-train a model. However, many different reactive datasets exist and combining this large amount of information could help  build a general transferable potentials for reactions in both condensed and gas phase without the need for millions of new QM calculations. Additionally, datasets have been created for many different combinations of elements. Meta-learning techniques could help build more transferable MLIPs over a wider range of elements with fewer calculations required. 

%An additional area, that is not explored in this works, is creating potentials that can move to new elements that have not previously been included. Meta-learning techniques could provide a way to incorporate information from different elements, and learn a representation that can adapt to any element present. 

However, combining multiple datasets together and training with meta-learning will not always improve results. This was seen with the CCSD(T) results where fitting straight from ANI-1x to CCSD(T) resulted in the lowest error. Therefore, adding more data when there is a specific application in mind is not always the best approach, particularly if the additional data is far from the final application. For specific applications, transfer learning from one dataset to another may yield the best training and test set errors. However, if multiple data sets need to be incorporate together, or a general model is desired which can be specialized to multiple different tasks, meta-learning methods are preferable.

%Problems exposed
With the techniques described in this work, multiple datasets can be fit at once. However, this advancement has exposed a more practical problem with the datasets currently published. There is not a standard format for storing information. Manual manipulation of datasets to a standard format is extremely time-consuming. The need for uniformity in the structure of datasets produced is therefore becoming increasingly important.

% Datasets, big data, pre-training conclude
 
The growth of available datasets containing quantum mechanical information for molecular and material structures has given researchers unprecedented levels of QM information. However, combining data from multiple data-sources is a major challenge. We have shown how meta-learning can be used to combine information from multiple datasets generated with varying levels of theory. This advancement changes the way that existing datasets should be viewed, and opens up new avenues for MLIP fitting. Beyond this, the results suggest that meta-learning can be seen as a general approach for combining training datasets for the broad array of chemical and materials processes where data science models can benefit.

\acknowledgments
This work was supported by the United States Department of Energy (US DOE), Office of Science, Basic Energy Sciences, Chemical Sciences, Geosciences, and Biosciences Division under Triad National Security, LLC (‘Triad’) contract grant no. 89233218CNA000001 (FWP: LANLE3F2). A. E. A. Allen and S. Matin also acknowledge the Center for Nonlinear Studies. Computer time was provided by the CCS-7 Darwin cluster at LANL.

\bibliography{refs}

\end{document}

\clearpage
\maketitle
\tableofcontents

\clearpage

\section{Supplementary Methods}

\subsection{Uncertainity of ANI-1x across Datasets}
\begin{figure}[h]
  \begin{center}
    \includegraphics[width=0.40\textwidth]{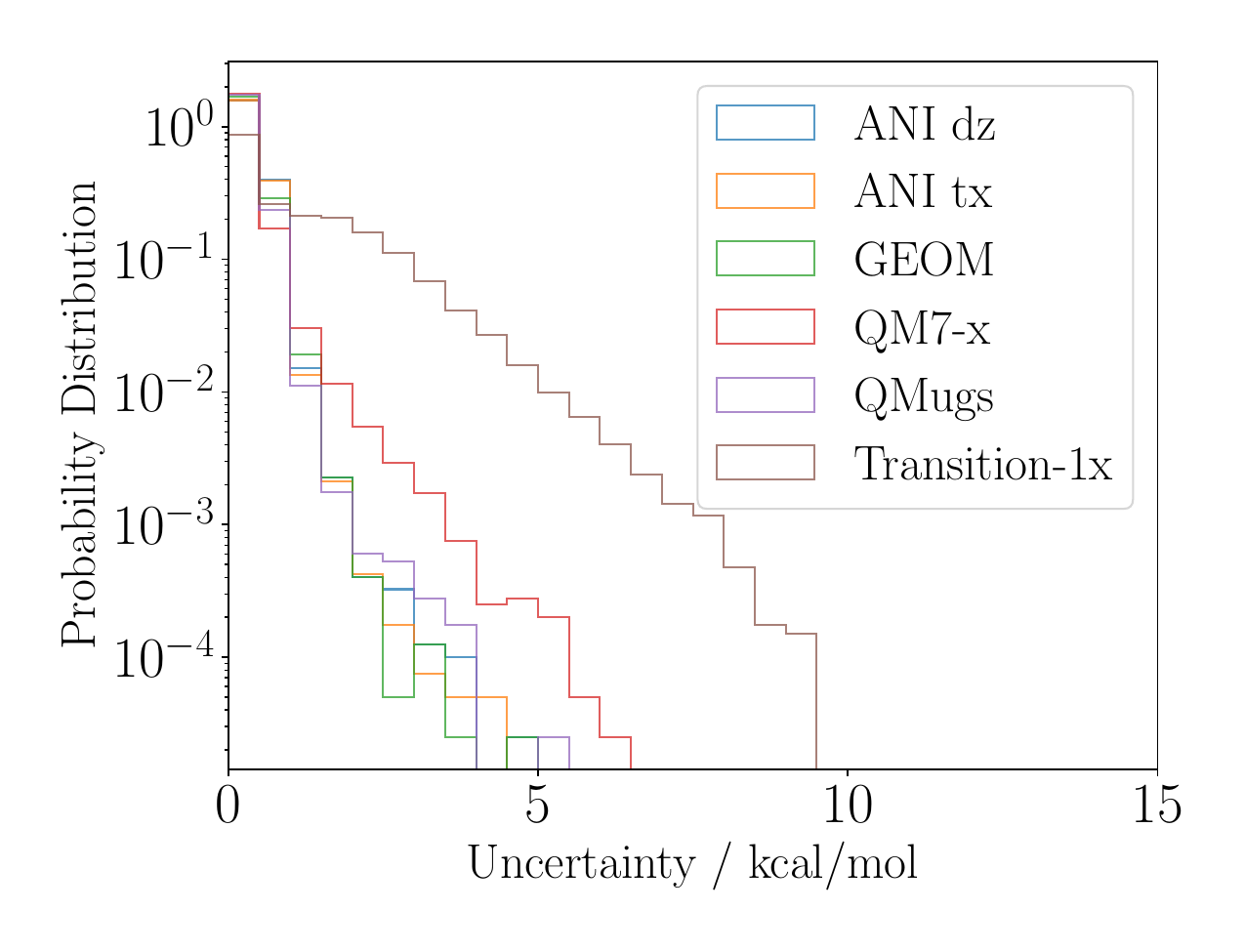}
  \end{center}
  \vspace{-0.75cm}
  \caption{ The distribution of uncertainties with the ANI-1x potential for 100,000 structures taken from the different datasets included in this work.  }
  \label{fig:uncert_dis}
\end{figure}

\subsection{Hyperparameter Optimization}
The meta-learning hyperparameters which control:
\begin{itemize}
    \item The number of steps taken at each optimization step, $k$, 
    \item How the parameters are updated given from the task individual NN parameters $\epsilon$ 
    \item The maximum number of epochs ($N_{max}$) for fitting to the final dataset
\end{itemize}    
were optimized. These are hyperparameters which specific to the Reptile meta-learning algorithm and not the molecular potential itself. 

\begin{figure}[h]
  \begin{center}
    \includegraphics[width=0.4950\textwidth]{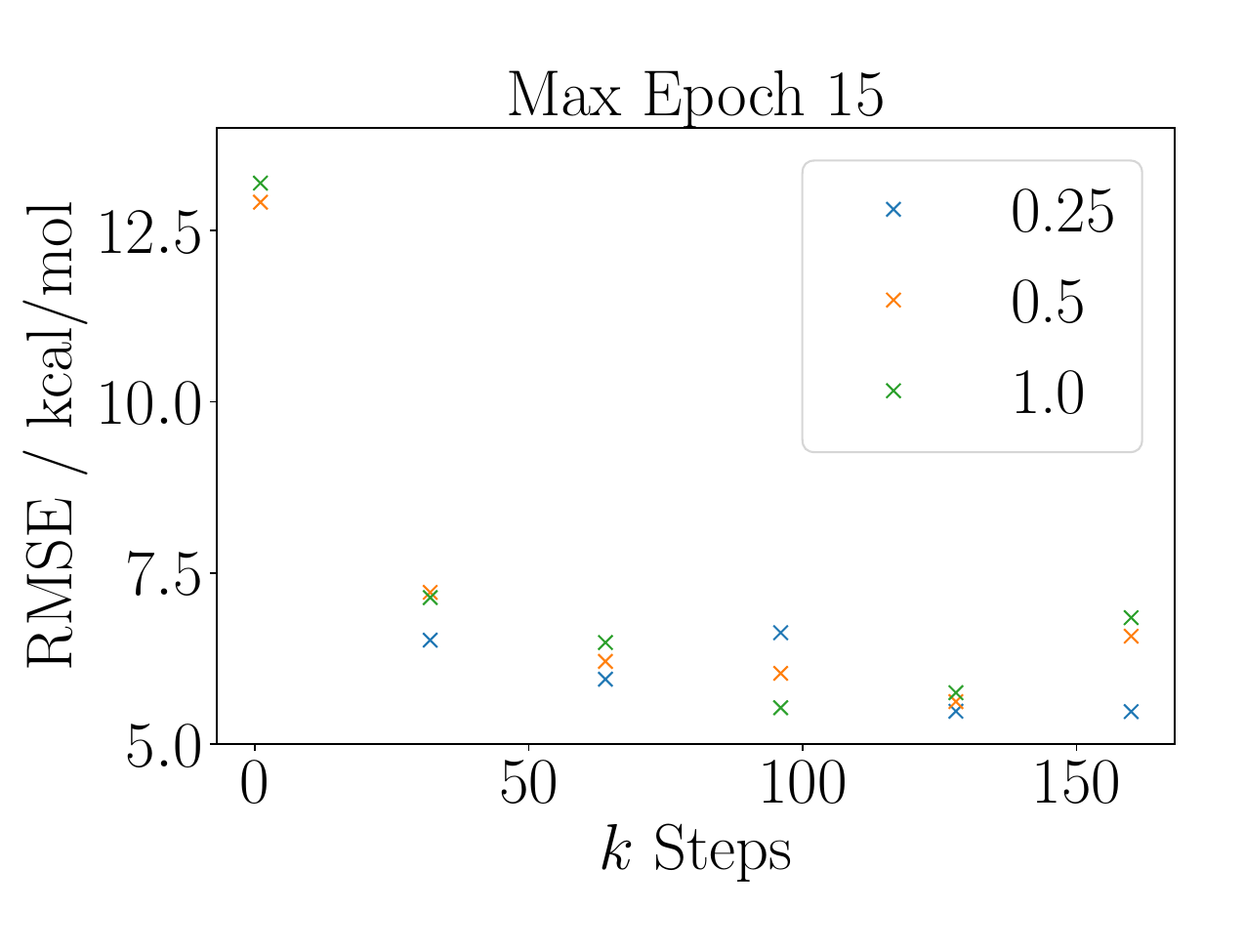}
    \includegraphics[width=0.4950\textwidth]{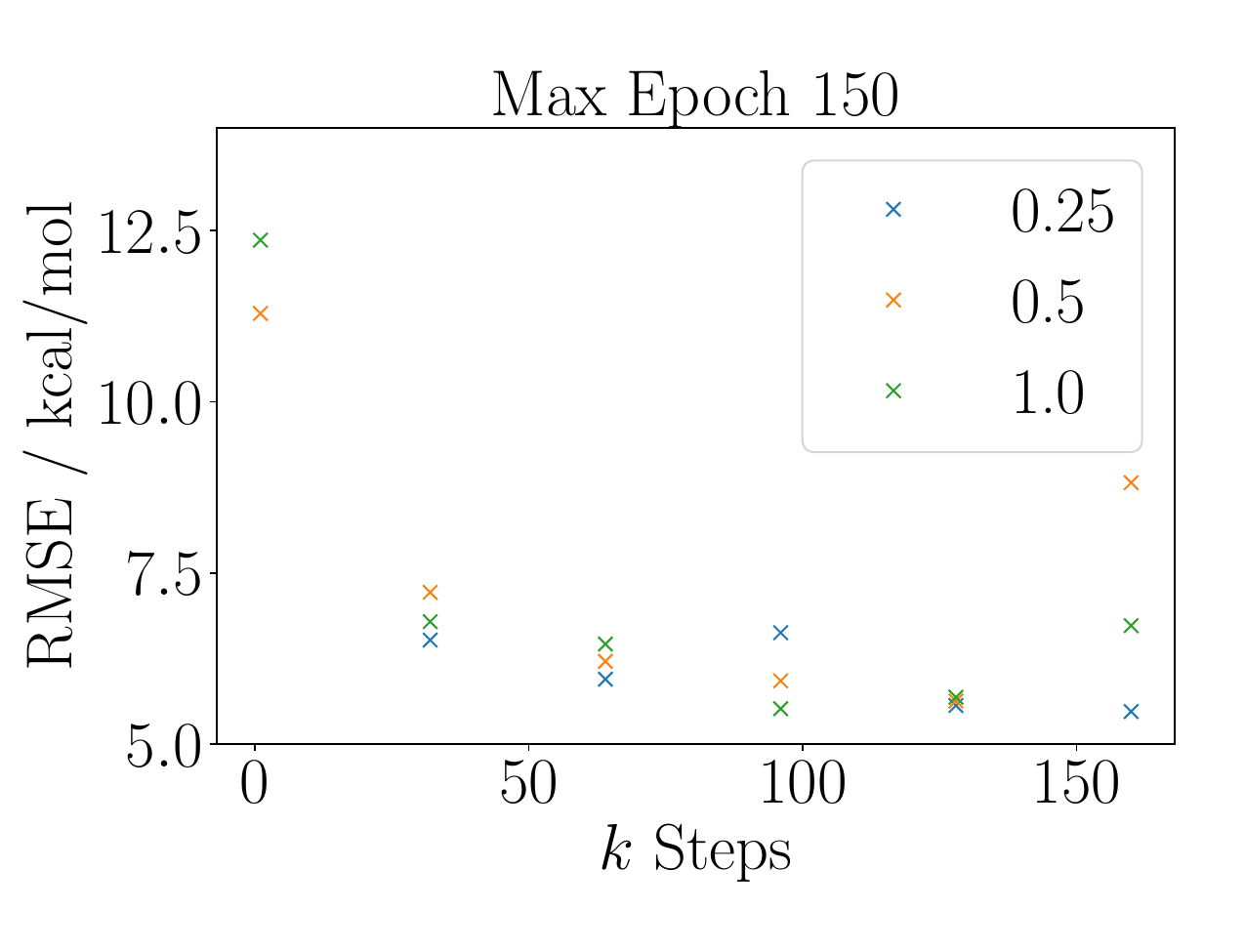}
    \includegraphics[width=0.4950\textwidth]{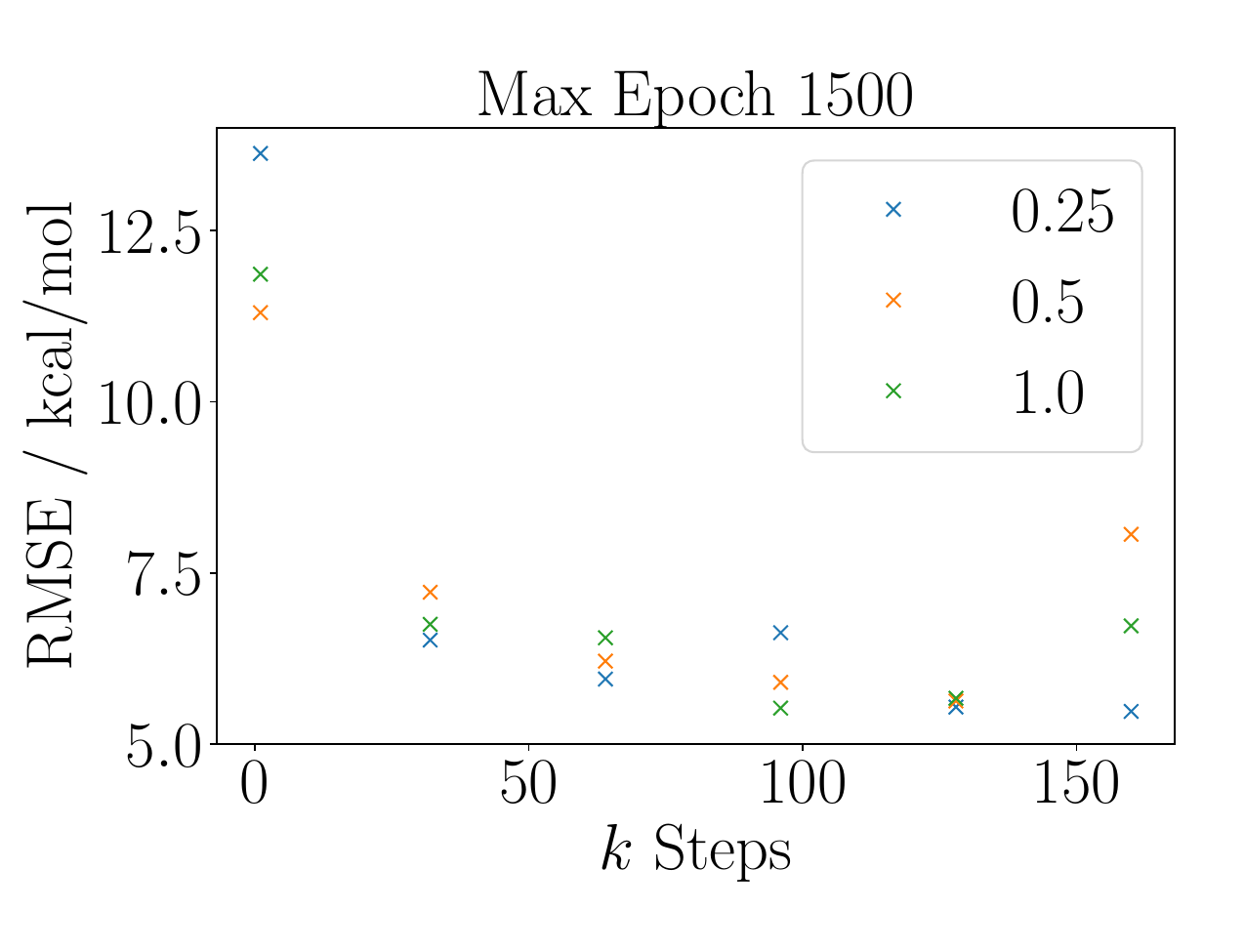}
    \includegraphics[width=0.4950\textwidth]{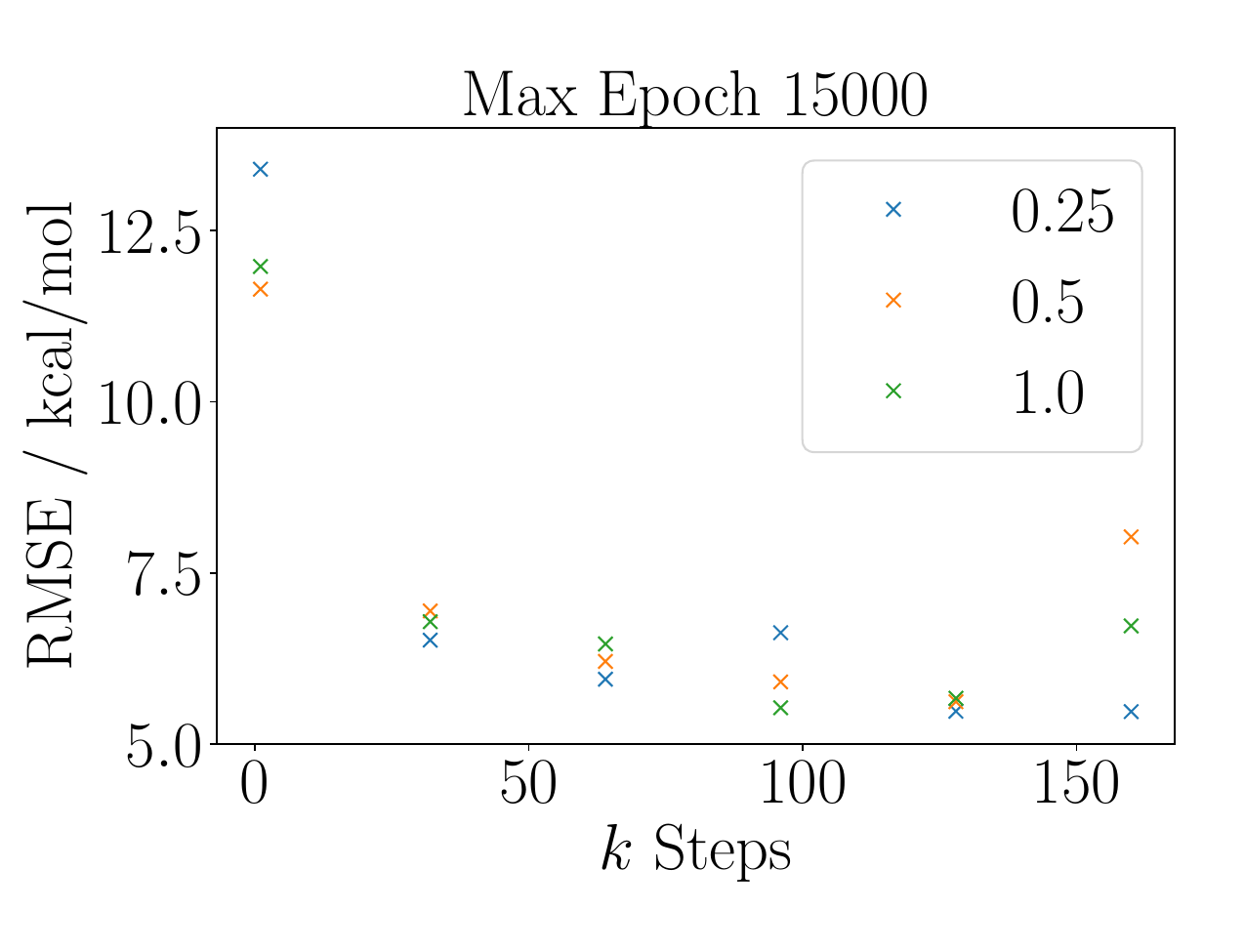}
  \end{center}
  \caption{The change in RMSE for the CCSD(T) test set for varying meta-learning hyperparameters. By 15,000 epochs the training has fully converged. If no training to DFT is performed and the potential is fit directly to the CCSD(T) data the error is 13.90 kcal/mol.   }
  \label{fig:HP}
\end{figure}

The change in the test set error with $k$, $\epsilon$ and $N_{max}$ is shown in Fig.~\ref{fig:HP}. These test were performed on a subset of 10,000 structures from each DFT dataset and 7,650 structures from the CCSD(T) dataset. The number of steps chosen (32,64,96) correspond to the number of complete passes through training data. This batch size used is 250. The optimal hyperparameters for the Reptile algorithm are $k=128$ and $\epsilon=0.25$, however very similar performance is seen for $k=96$ and $\epsilon=1.00$ and as fewer $k$ steps results in faster convergence this was chosen as the optimal value.  The value of $k$ used in this work was increased as the dataset size increases so that it consistently corresponds to 3 complete passes through each dataset. The performance does not change as the maximum number of epochs are increased and the CCSD(T) training tends to converge within 150 epochs. The default retraining used for the meta-learning potential throughout this work is 150 epochs.

\begin{table}[]
\begin{tabular}{r|c|}
\cline{2-2}
 & \textbf{\begin{tabular}[c]{@{}c@{}}CCSD(T) Error\\ kcal/mol\end{tabular}} \\ \hline
\multicolumn{1}{|r|}{\textbf{AN1-1x dx}} & 6.44 \\ \hline
\multicolumn{1}{|r|}{\textbf{ANI-1x tz}} & 6.98 \\ \hline
\multicolumn{1}{|r|}{\textbf{GEOM}} & 16.46 \\ \hline
\multicolumn{1}{|r|}{\textbf{Transition1x}} & 11.73 \\ \hline
\multicolumn{1}{|r|}{\textbf{QM7-x}} & 8.26 \\ \hline
\multicolumn{1}{|r|}{\textbf{Qmugs}} & 14.69 \\ \hline
\multicolumn{1}{|r|}{\textbf{ANI-1ccx}} & 13.9 \\ \hline
\end{tabular}
\caption{The error for the CCSD(T) data after first fitting the six datasets. The ANI-1ccx result is when no previous training is carried out. }
\label{tab:ind_err}.
\end{table}

The CCSD(T) error from individual transfer learning from data-sets are shown in Table~\ref{tab:ind_err}. The ANI-1x datasets produce the best results, this is to be expected as they contain the ANI-1ccx configurations. GEOM and QMugs do not help fitting, this may be because both only contain energies and not forces. An improvement of around 20\% for the CCSD(T) be seen when multiple datasets are used with 10,000 datapoints from each dataset.   

\clearpage

\subsection{Stages of fitting}
\begin{figure}[h]
  \begin{center}
    \includegraphics[width=0.75\textwidth]{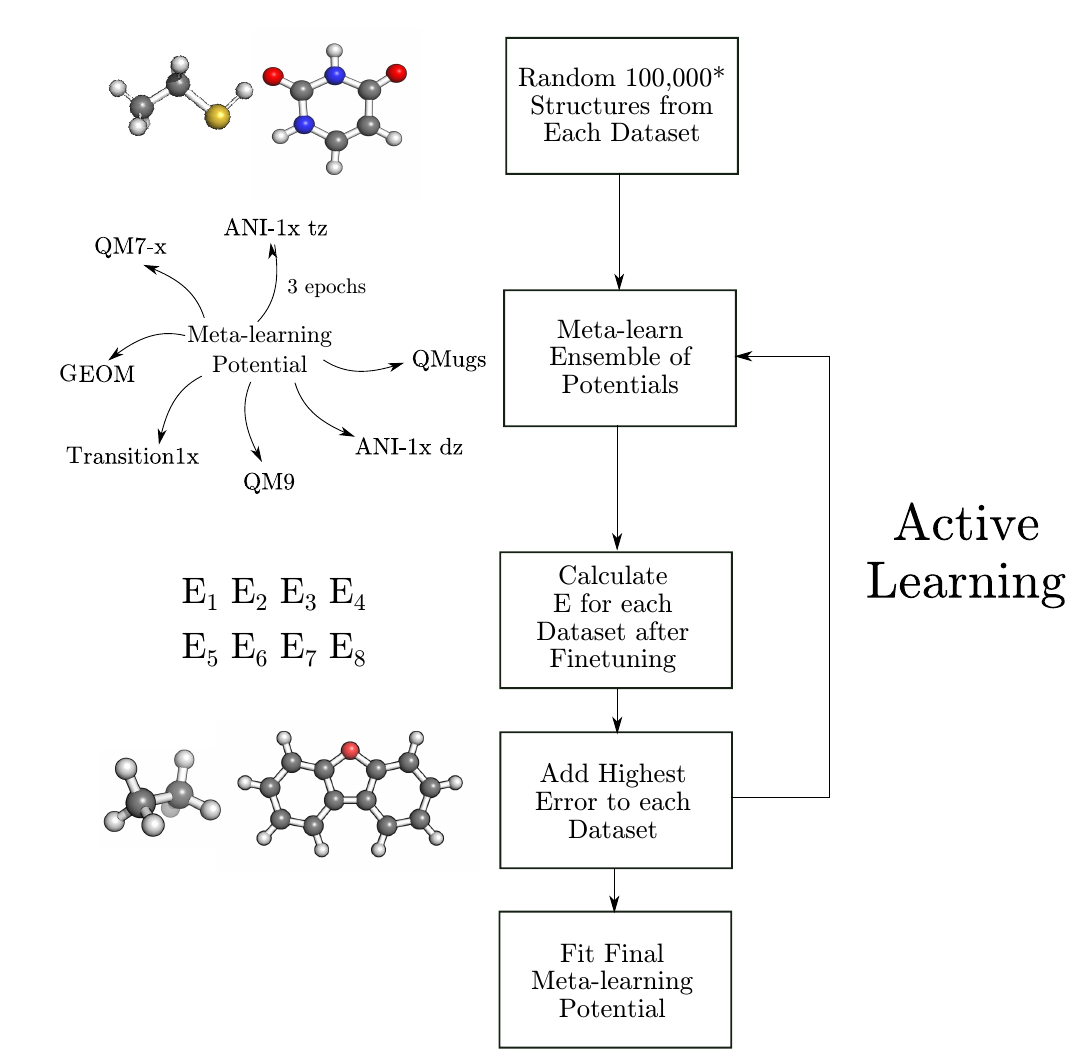}
  \end{center}
  \caption{The stages of fitting in the meta-learning process. *The QM9 is the exception with only 10,000 structures. }
  \label{fig:stages_fitting}
\end{figure}

\subsection{QM9 Weighting}
The QM9 dataset contains the configurations of 133,885 organic molecules. The energy of these configurations was calculated with 76 different exchange-correlation functionals and 3 basis sets. Therefore, this dataset contains 228 separate tasks. If all of these tasks were included in every loop of the Reptile algorithm this would heavily bias the potential to the equilibrium QM9 structures. To overcome this, we added a weighting parameter $w$. At every loop of the Reptile cycle a task is included with probability $p=1/k$. This allows over-represented tasks to be included less frequently. 

The weighting for the QM9 tasks were set to 228, all other datasets had a weighting of 1 - meaning that they were always included in the fitting process. 

\subsection{Validation and Learning Rate Schedule}

As multiple datasets are being fit at once, there are validation errors measured for each dataset. These errors are calculated after $k$ steps have been taken towards the dataset. The validation error therefore follows the error after `fine-tuning'. 

The learning rate schedule starts at 0.001 and decreases by 50$\%$ after 150 epochs have not produced a decrease in error. This continues until the learning rate reaches 0.00001, at this stage the training is stopped. 

\subsection{Adding New Structures}

For iteration 1/2/3 the following criteria is used: 
\begin{itemize}
    \item For QMugs, ANI-1x-dz, ANI-1x-tz, GEOM and QM7-x structures with an error greater than 3.5/2.5/1.5 kcal/mol per atom were added to the datasets. 
    \item For Transition-1x, the structures are highly correlated as they all originate an approximation to the minimum energy path. Therefore, structures were separated into their reaction paths. The five highest error structures for each path were then added only if the energy differs from the existing structures added in that path by at least 0.1 kcal/mol.  
\end{itemize}

The final number of structures present are given in Fig.~\ref{tab:AL_DS}. 

\begin{table}[]
\begin{tabular}{r|c|}
\cline{2-2}
 & \textbf{Number of Structures} \\ \hline
\multicolumn{1}{|r|}{\textbf{ANI-1x dz}} & 855,028 \\ \hline
\multicolumn{1}{|r|}{\textbf{ANI-1x tz}} & 732,154 \\ \hline
\multicolumn{1}{|r|}{\textbf{Qmugs}} & 103,374 \\ \hline
\multicolumn{1}{|r|}{\textbf{QM7-x}} & 1,085,249 \\ \hline
\multicolumn{1}{|r|}{\textbf{GEOM}} & 122,552 \\ \hline
\multicolumn{1}{|r|}{\textbf{Transition-1x}} & 251,095 \\ \hline
\multicolumn{1}{|r|}{\textbf{Total}} & 3,149,452 \\ \hline
\end{tabular}
\caption{The number of structures included from each dataset. }
\label{tab:AL_DS}
\end{table}

\clearpage

\subsection{3BPA}

\subsubsection{Energy RMSE}

\begin{figure}[h]
  \begin{center}
    \includegraphics[width=0.49\textwidth]{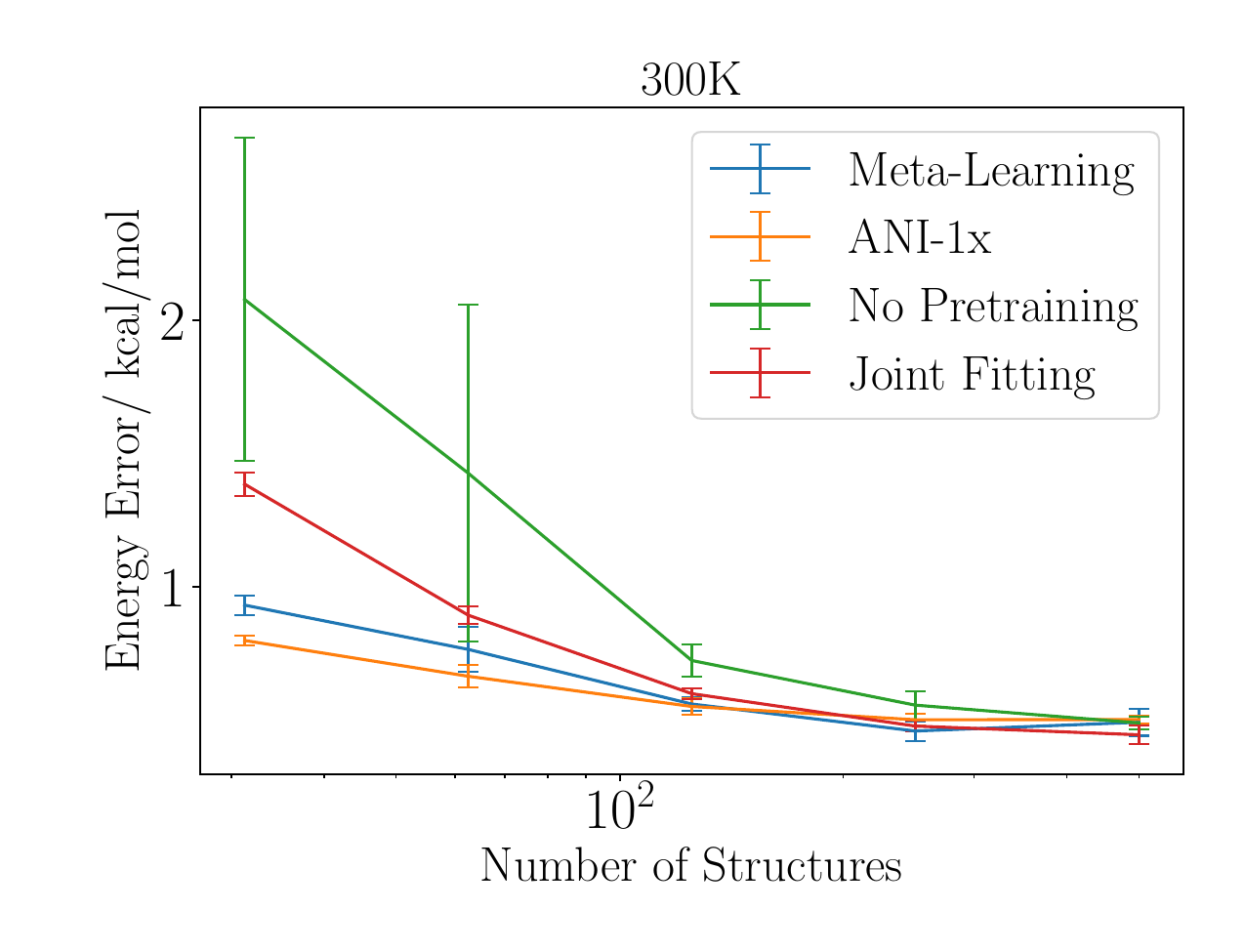}
    \includegraphics[width=0.49\textwidth]{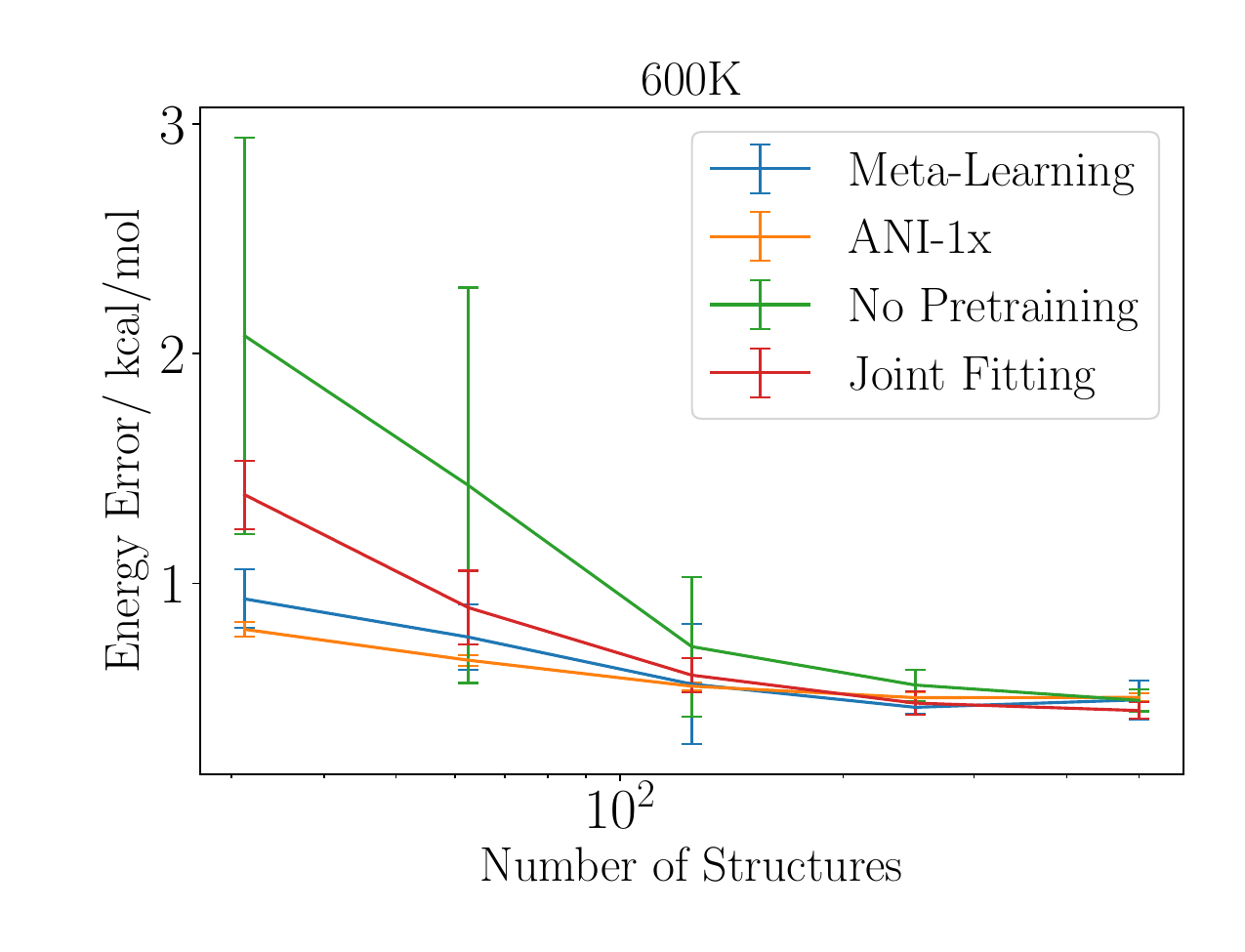}
    \includegraphics[width=0.49\textwidth]{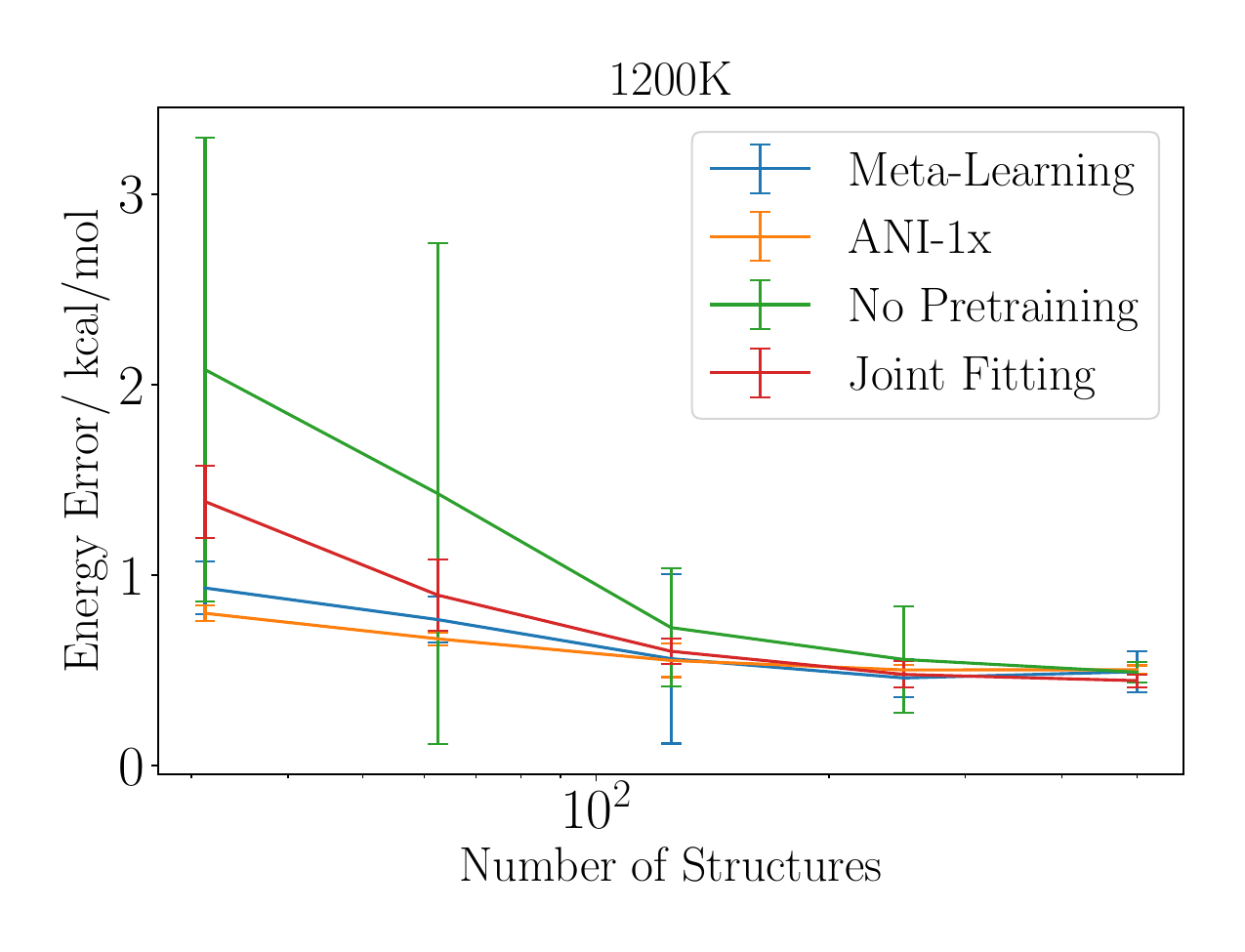}
  \end{center}
  \caption{The learning curve for the energy RMSE for the 3BPA molecule at three temperatures.  }
  \label{fig:DS_iter1}
\end{figure}

\clearpage

\subsubsection{Force RMSE}

\begin{figure}[h]
  \begin{center}
    \includegraphics[width=0.49\textwidth]{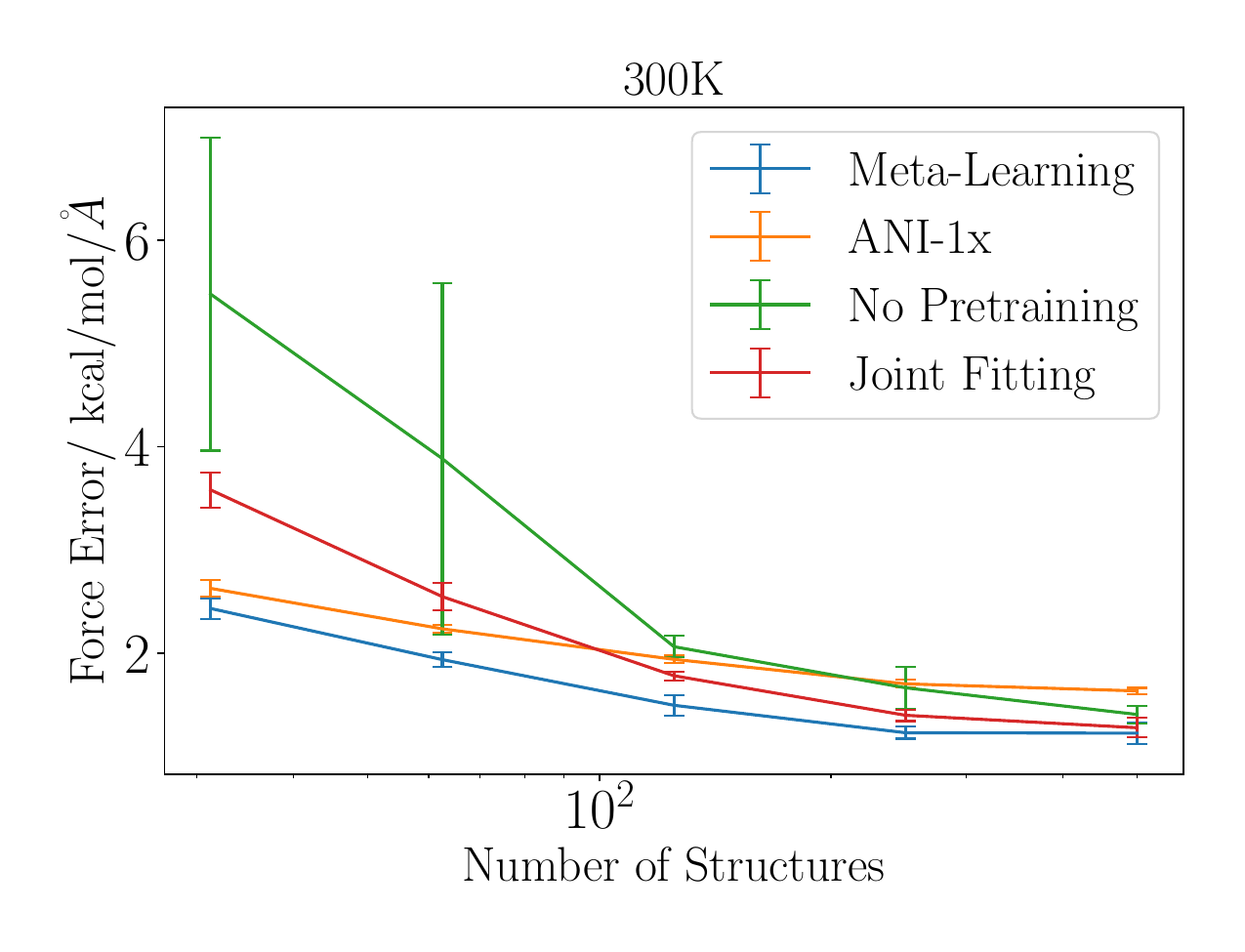}
    \includegraphics[width=0.49\textwidth]{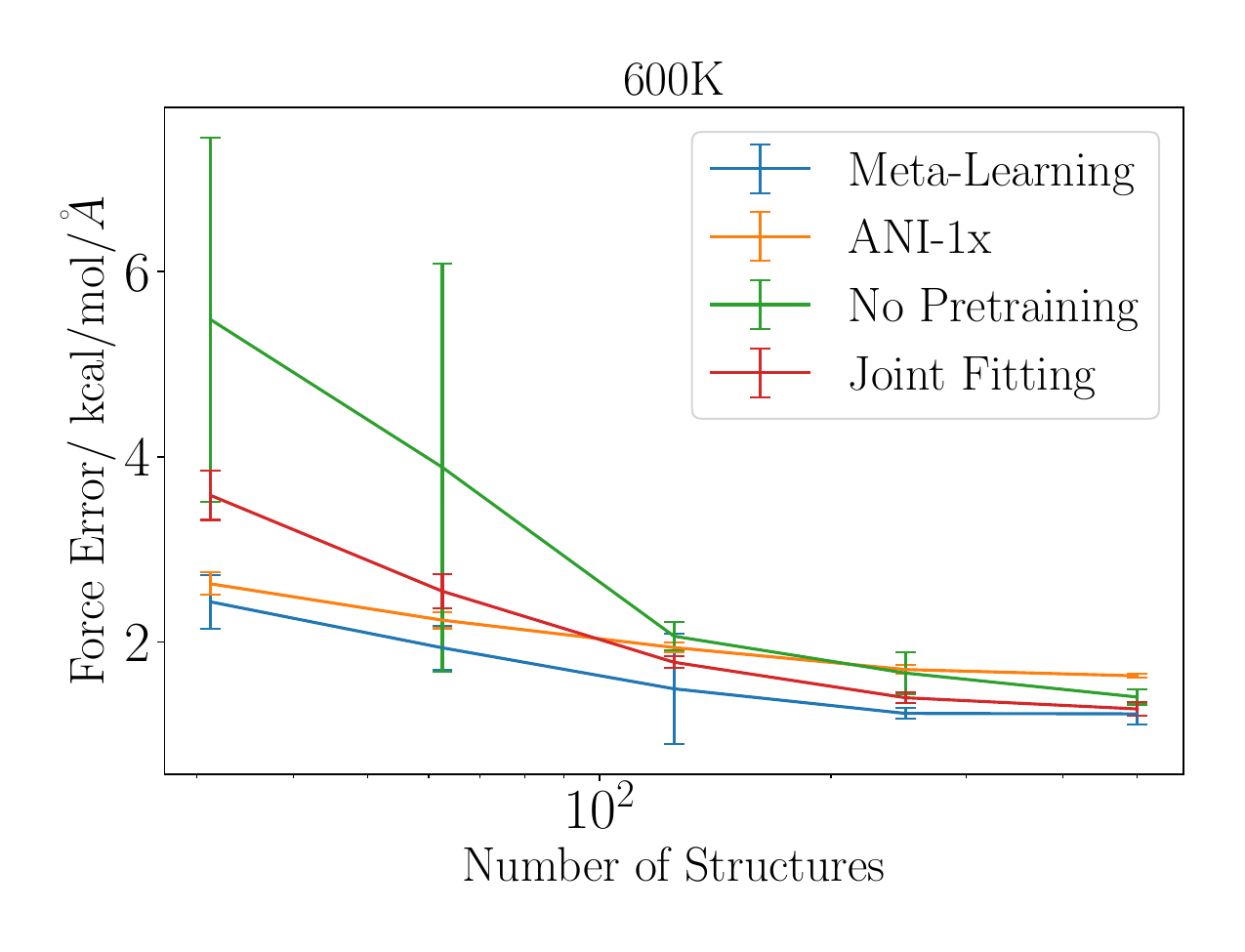}
    \includegraphics[width=0.49\textwidth]{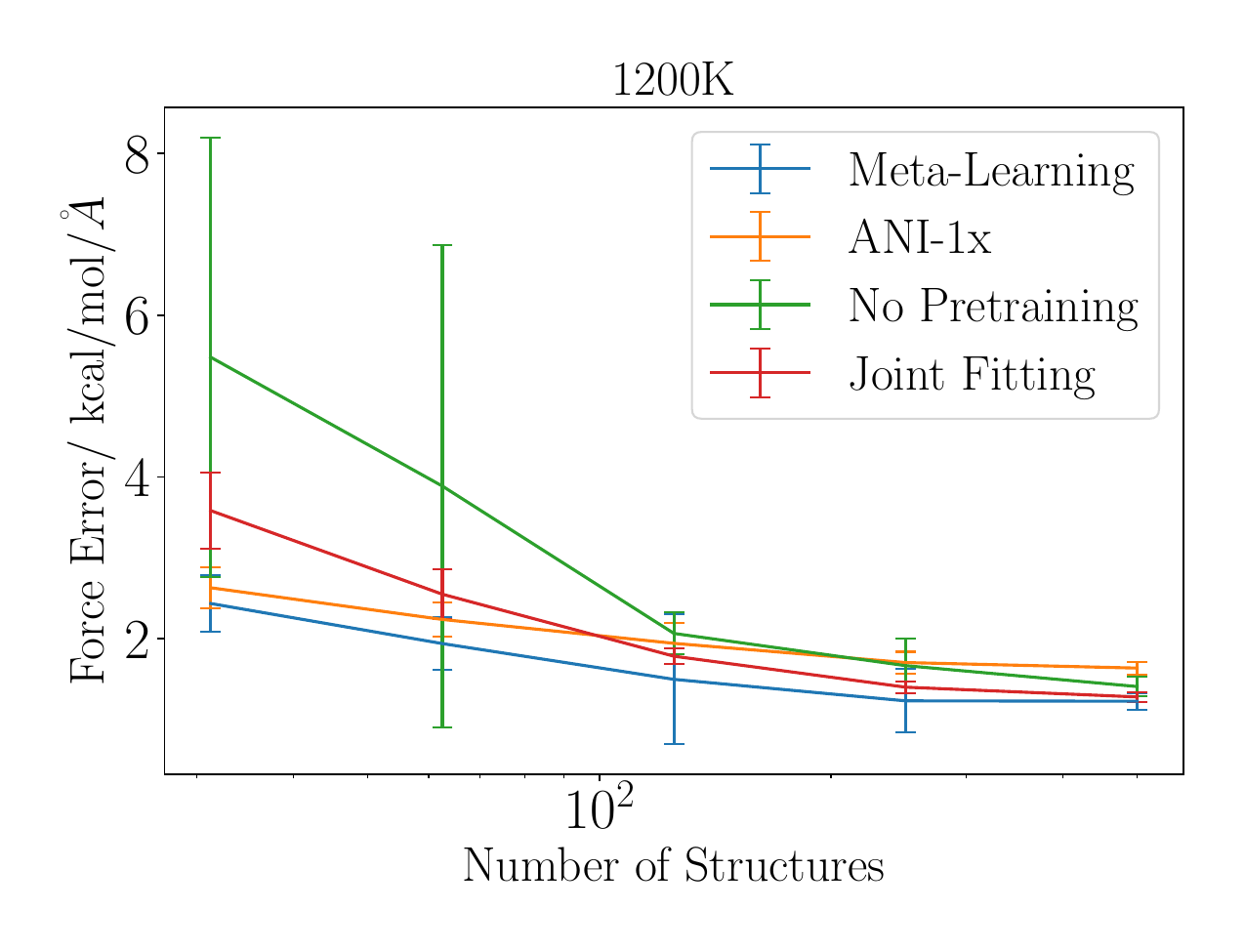}
  \end{center}
  \caption{The learning curve for the force RMSE for the 3BPA molecule at three temperatures. }
  \label{fig:DS_iter1}
\end{figure}

\clearpage

\subsubsection{Dihedral Scans}

\begin{figure}[h!]
  \begin{center}
    \includegraphics[width=0.495\textwidth]{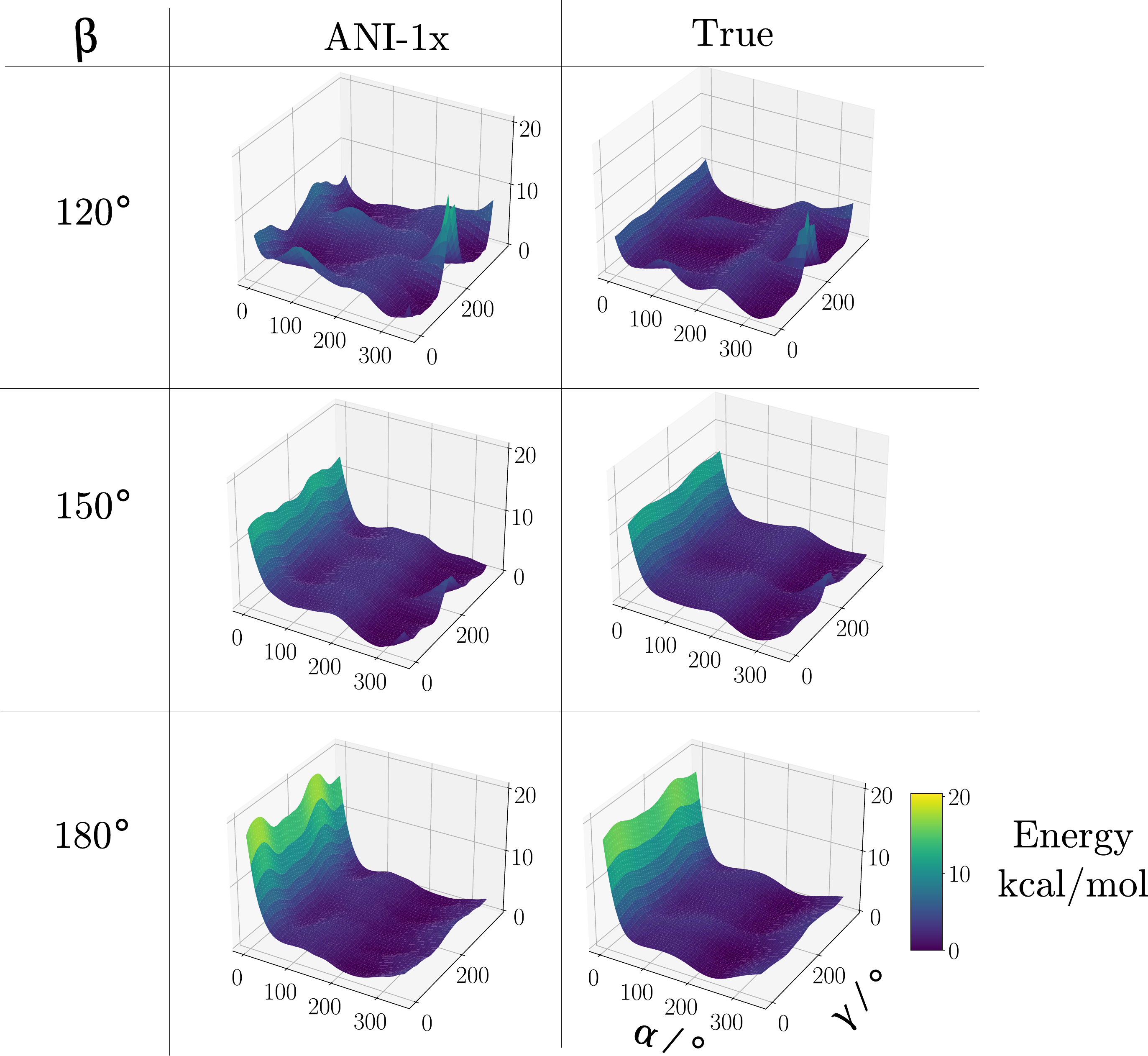}
  \end{center}
  \caption{The dihedral angle energy scans for the 3BPA molecule at three different $\beta$ angles for a potential pre-trained to ANI-1x and the QM results. }
  \label{fig:3BPA-H-bond-31}
\end{figure}

\subsubsection{Hydrogen Dissociation}

\begin{figure}[h!]
  \begin{center}
    \includegraphics[width=0.495\textwidth]{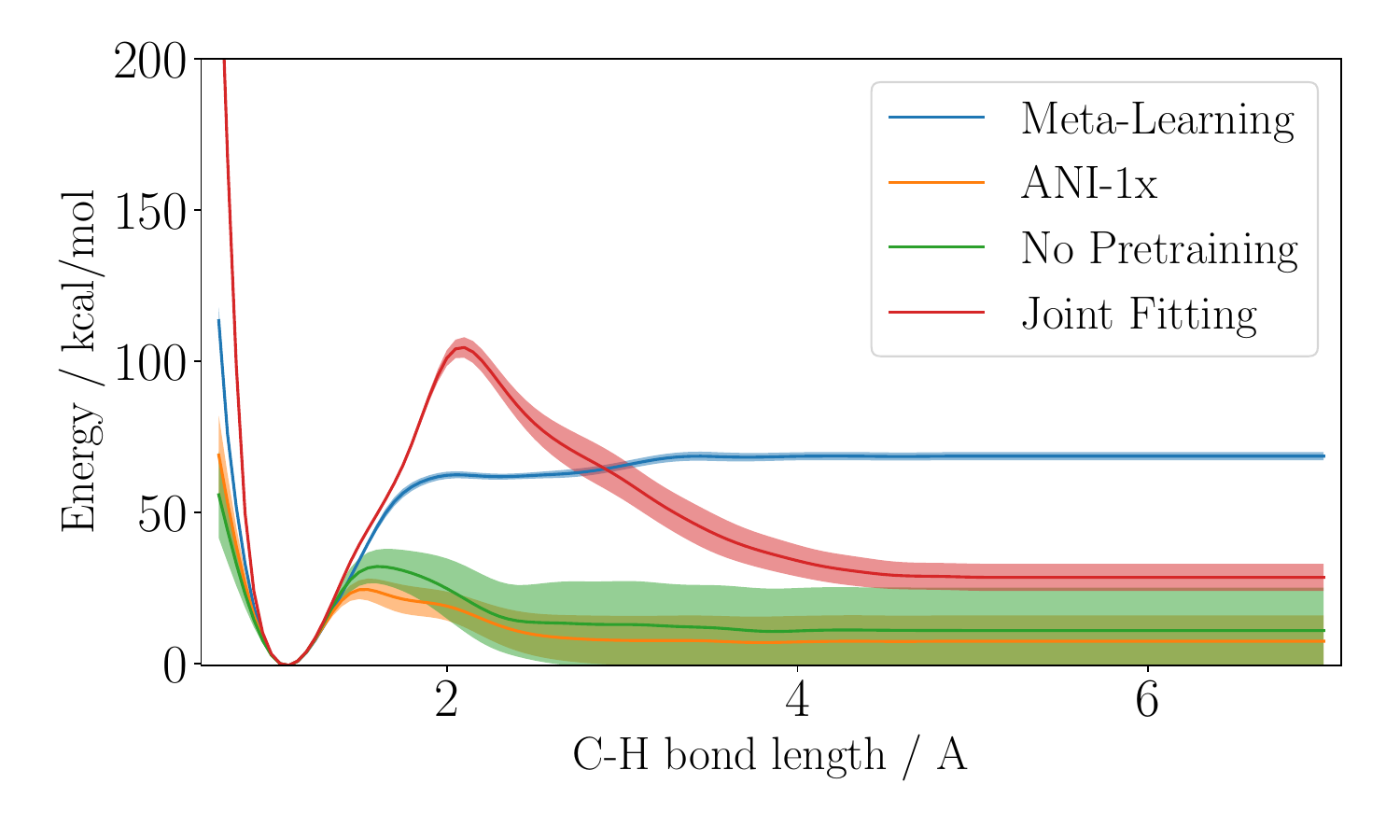}
  \end{center}
  \caption{The bond dissociation energy for the removal of a hydrogen atom when trained to 31 structures of 3BPA.}
  \label{fig:3BPA-H-bond-31}
\end{figure}

\begin{figure}[htb!]
  \begin{center}
    \includegraphics[width=0.45\textwidth]{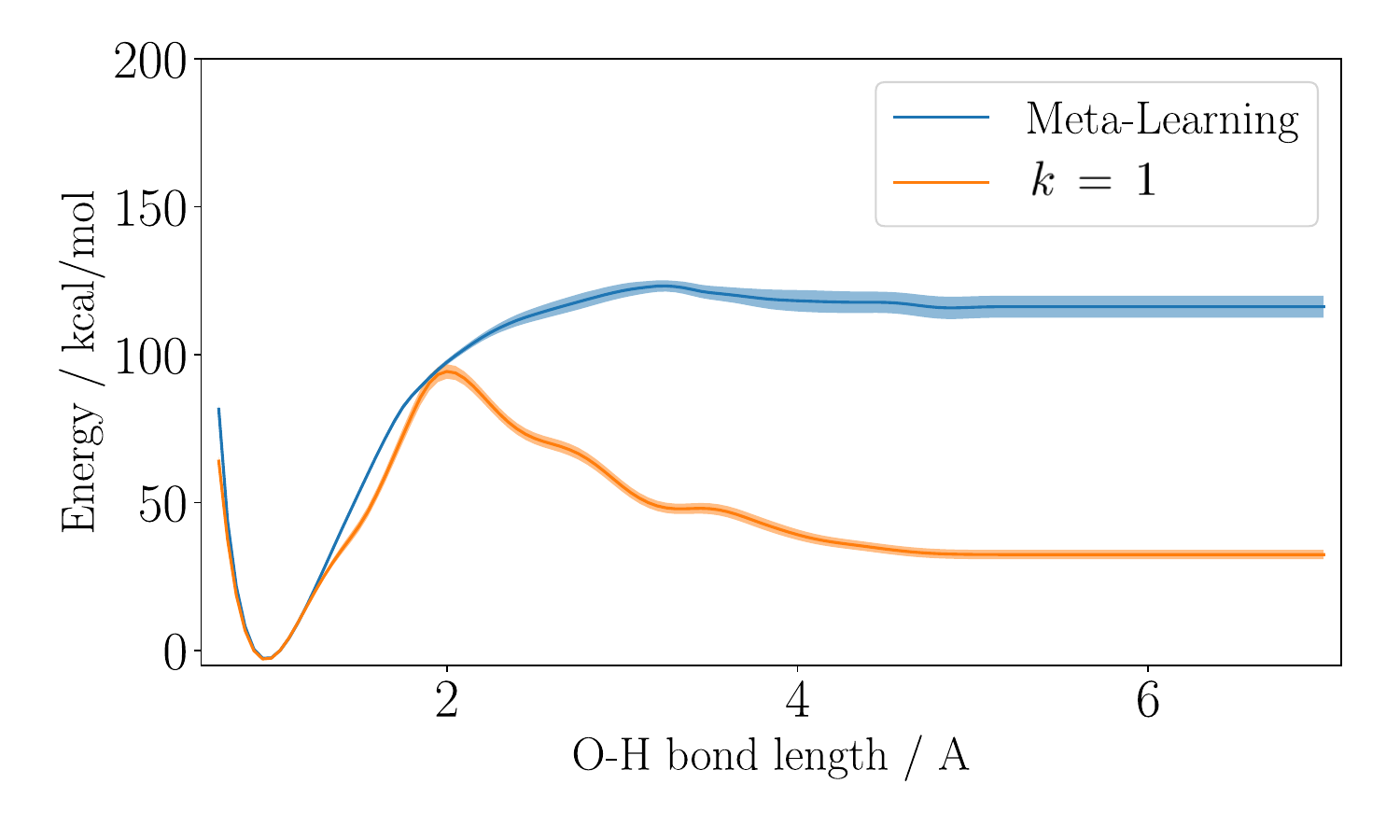}
  \end{center}
  \caption{The bond dissociation energy when trained to the rMD17 structures for an ethanol molecule. Again, smooth dissociation is not seen when joint fitting is used. The potential was refit for a maximum of 150 epochs. }
  \label{fig:ethanol_H_bond}
\end{figure}

\clearpage

\subsection{Fitting to QM9}

\begin{figure}[htb!]
  \begin{center}
    \includegraphics[width=0.45\textwidth]{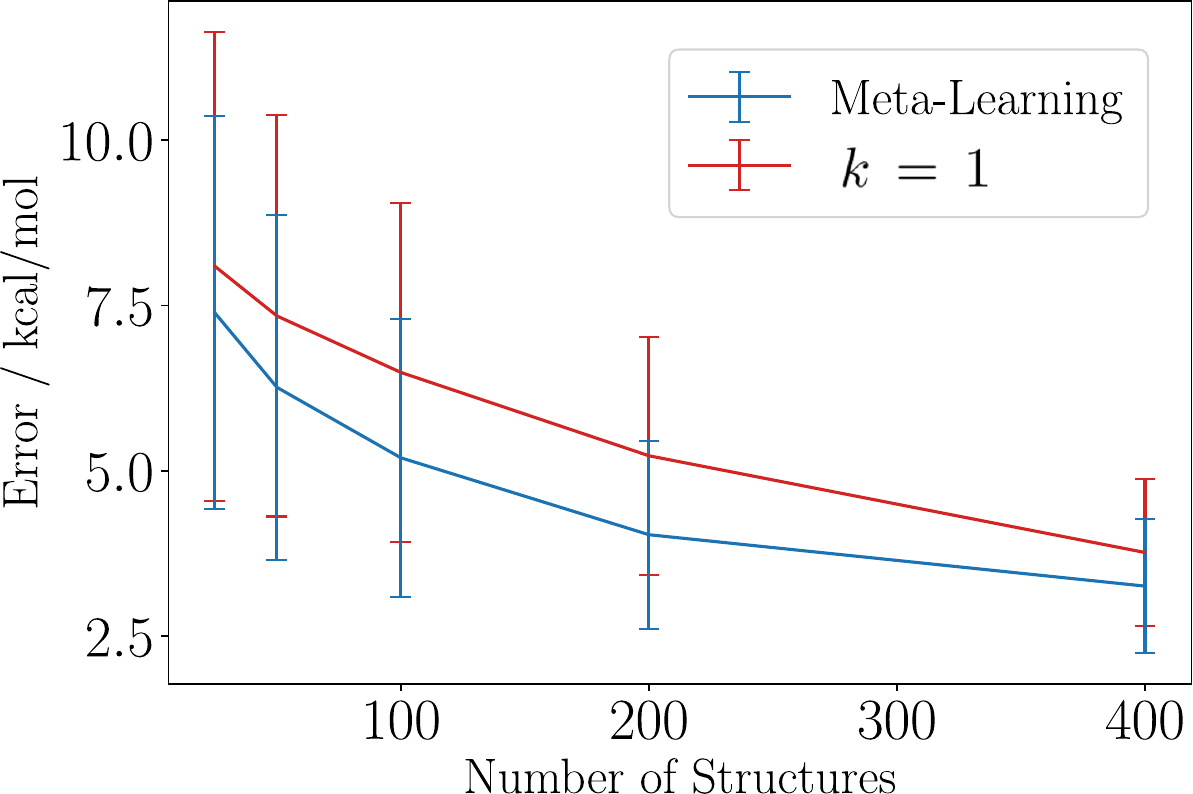}
  \end{center}
  \caption{The error as a function of the number of structures averaged across 5 different DFT functions and 3 basis sets, for different fitting procedures for the QM9 dataset.  The meta-learning algorithm fits the different levels of theory together with $k=10$ used and is compared to $k=1$. The standard deviation of the error bars corresponds to the variation across different levels of theory.   }
  \label{fig:QM9_b}
\end{figure}

%\bibliography{refs}